\documentclass[twocolumn]{aastex63}

\submitjournal{ApJ}

\usepackage{graphicx}
\usepackage{dcolumn}
\usepackage{bm}

\usepackage{makecell}
\usepackage{verbatim} 
\usepackage{amsmath}
\usepackage{comment}

\begin{document}



\title{The Effects of Kinematic MHD on the Atmospheric Circulation of Eccentric Hot Jupiters}

\correspondingauthor{Hayley Beltz}
 \email{hbeltz@umd.edu}

\author[0000-0002-6980-052X]{Hayley Beltz}

 \affiliation{Department of Astronomy, University of Maryland, College Park, MD 20742, USA}

 \author[0009-0003-5661-639X]{Willow Houck}
\affiliation{Department of Astronomy, University of Maryland, College Park, MD 20742, USA}

\author[0000-0002-4321-4581]{L. C. Mayorga}
\affiliation{Johns Hopkins Applied Physics Laboratory, Laurel, MD, 20723}
  \author[0000-0002-9258-5311]{Thaddeus D. Komacek}
\affiliation{Department of Astronomy, University of Maryland, College Park, MD 20742, USA}

\author[0000-0003-3888-3753]{Joseph R. Livesey}
    \affiliation{Department of Astronomy, University of Wisconsin--Madison, Madison, WI 53706}
    
\author[0000-0002-7733-4522]{Juliette Becker}
   \affiliation{Department of Astronomy, University of Wisconsin--Madison, Madison, WI 53706}

\begin{abstract}
Hot Jupiters are typically considered to be tidally locked due to their short orbital periods. The extreme irradiation can result in atmospheric species becoming thermally ionized on the dayside, which then interact with the planet's magnetic field by resisting flow across magnetic field lines, shaping the atmospheric structure. However, an eccentric orbit results in temporally dependent irradiation and a non-permanent dayside, as the planet-star distance can change drastically during its orbit. In this paper, we present 3D atmospheric models of TOI-150b, an eccentric (e=0.26), Jupiter-mass ($\sim 1.75 M_{\mathrm{Jup}}$) planet whose equilibrium temperature varies from 1300K to 1700K. We conduct simulations for magnetic field strengths ranging from 0-30 Gauss using the kinematic magnetohydrodynamics (MHD) approach. When compared to simulations of the planet assuming a circular orbit, we find that the eccentric orbit results in a strengthened and narrowed equatorial jet, westward winds at mid-latitudes, and a phase-dependent thermal inversion. The strength and magnitude of these effects scale with the chosen global magnetic field strength. We also generate high-resolution ($R=100,000$) emission spectra to study net Doppler shifts and find inter-orbit spectroscopic variability at moderate magnetic field strengths, as well as decreased Doppler broadening as magnetic field strengths increase. This work represents the first time that the kinematic MHD approach has been applied to an eccentric hot Jupiter and highlights the importance of a locally calculated, temperature dependent magnetic drag prescription for predicting atmospheric structure and resulting spectra. \\

\end{abstract}

\section{Introduction}
Hot Jupiters (HJs) are typically characterized by their large masses and short orbital periods. In most cases, these orbits are circular and assumed to be tidally spin-synchronous, resulting in a permanent dayside and a permanent nightside. Extensive theoretical work has focused on modeling the atmospheres of hot Jupiters in 1D and 3D. By comparing these models with observations, a standard picture of the atmospheric circulation of hot Jupiters on  circular orbits has emerged \citep{Showman:2020rev}. The first 3D atmospheric model for a hot Jupiter by \cite{Showman2002} predicted the presence of a broad superrotating equatorial jet. This feature is ubiquitous in atmospheric models of hot Jupiters and causes the hottest part of the planet to be shifted eastward of the substellar point (\citealp{2007Knutson, 2009Menou, 2010Crossfield, 2013DobbsDixon, 2018Koll, 2018Mendonca}). This jet is driven by the combination of slow rotation and large temperature contrasts between the dayside and nightside of the planet, and is expected to increase in speed as the temperature contrast increases (\citealp{Showman2002, 2011ShowmanPolvani,  2017Komacek}).
These features are present across multiple general circulation models (GCMs)---multidimensional numerical models that simulate the atmospheric dynamics of planets---including the Unified Model, the SPARC/MITgcm, and the THOR GCM, among others \citep{MARLEY1999,2003Cho, Adcroft2004, 2009Menou, 2009Showman, 2012DobbsDixon, Mayne2014, 2016Joao}.

However, there exists a subset of objects on significantly eccentric orbits, such as HD 80606b (e=0.93), HAT-P-2b (e=0.52), and XO-3b (e=0.26) \citep{2001Naef, 2008Loeillet, 2008JohnsKrull}. These planets have stellar irradiation changing in time, which influences the planet's atmospheric circulation and chemistry. In addition, once the orbit is no longer circular, we no longer expect synchronous rotation, meaning that there is no longer a constant dayside and nightside. Previous work suggests these objects may exhibit pseudo-synchronous rotation, where the 
same side of the planet faces the star at periastron \citep{Hut1981}. This faster pseudo-synchronous rotation is expected to change the atmospheric circulation pattern \citep{2008LangtonA, 2008LangtonB, 2010Lewis, Kataria2013, 2014Lewis}. 

A wide range of previous theoretical work has explored the impact of eccentricity on hot Jupiter atmospheres. 
\cite{Kataria2013} studied eccentric hot Jupiters using the SPARC/MITgcm with equilibrium temperatures varying from 476K-1199K\footnote{Equilibrium temperatures were calculated based on the orbit-averaged stellar fluxes.} and eccentricities ranging from 0.0 to 0.75. \cite{Kataria2013} confirmed that eccentric hot Jupiters share many of the same global atmospheric features as circular planets, but are still significantly influenced by the introduction of eccentricity. When the eccentricity increased, the equatorial jet narrowed and strengthened, mid-latitude westward jets developed, and the day-night temperature contrast increased. They also found that the light curves were strongly dependent on the viewing geometry of the system. 
Furthermore, \cite{2014Lewis} simulated the atmospheric dynamics of the hot Jupiter HAT-P-2b, which has an eccentricity of e=0.5, using the SPARC/MITgcm. They found that inclusion of TiO/VO induces dayside thermal inversions that cause an increase in the global day-night temperature gradient. 
\cite{2021Mayorga} looked at three eccentric hot Jupiters with eccentricities equal to or greater than e=0.517 using a 1D cloudless model to study atmospheric dynamics and star-to-planet flux ratios. \cite{2021Mayorga} predicted that high eccentricity would cause an offset in the peak flux. This offset was consistently positive, meaning the peak flux occured after periastron. They suggested that this meant there would be a rapid increase in heating as the planet approaches periastron, with a slower decrease post-periastron.
Finally, \cite{2023Tsai} studied how eccentricity impacts the chemistry of hot Jupiters, specifically HD-80606b, and found that changes in temperature from varying distance to the host star are linked to transient changes in chemistry near periastron. This work used the models from \cite{Lewis2017}, which found that rotation period strongly impacts the timing of the planetary flux at periapse, with an assumed rotation period of twice the pseudo-synchronous value best reproducing \textit{Spitzer} observations.


Magnetic drag is expected to shape the atmospheric wind and thermal structures of HJs once they reach $\sim 1500~\mathrm{K}$, due to the increased ionization of the atmosphere \citep[][]{Perna2010magdrag,2012RauscherGCM, Rogers_2014b, 2021Hindle2021hotspotreversals, Beltz2022a}. Previous non-ideal MHD modeling work has demonstrated increased atmospheric variability for ultra-hot Jupiters, particularly near the equator \citep{Rogers2017}. This can result in a time-varying hotspot offset which can reverse from eastward to westward in the case of atmospheres that are hot enough to host an atmospheric dynamo  \citep{Rogers2017McElwain}. Higher internal dipole field strengths have a greater impact on the hotspot offset. 

However, the full non-ideal MHD equations have never been coupled to a GCM with realistic radiative transfer. GCMs often use a uniform drag timescale to capture the effects of magnetic drag \citep{PerezBecker2013,2015Kataria, Komacek2016, 2018Koll, Kreidberg2018, 2018Mansfield, 2019Arcangeli}. This approach uses a single timescale, applied numerically as a Rayleigh drag\footnote{Despite its common use in GCMs, applying a Rayleigh drag can result in a loss of momentum that is not conserved. See \cite{Shaw2007} for a more thorough discussion},  to represent magnetic effects (or other sources of frictional drag, for instance shear instabilities, \citealp{2016Fromang}) globally. While this can reproduce some of the expected effects of magnetism (e.g slower wind speeds and small to no hotspot offset), this approximation can be problematic as it implicitly assumes magnetic fields that are orders of magnitude stronger on the nightside than dayside \citep{Beltz2022a}. For a more physically motivated approximation of magnetic effects, \cite{Perna2010magdrag} introduced a locally calculated magnetic drag timescale. This timescale is a function of magnetic resistivity---and thus temperature---allowing the strength of the drag to vary in a more realistic, consistent manner. This timescale was first implemented in \cite{RauscherMenou2013} on two hot Jupiters and in \cite{Beltz2022a} on an ultra-hot Jupiter, and is ultimately what is used in this work. We describe this method further in Section \ref{subsec: GCM}.


The work we present in this study is novel in that it is the first time a GCM has been used to study magnetic effects in a the atmosphere of an \textit{eccentric} hot Jupiter. Due to the changing irradiation, the temperature structure of the planet will vary more than for a circular case, further highlighting the importance of using a magnetic drag timescale that is calculated based on local conditions. 
By applying magnetic drag and studying how eccentricity impacts these atmospheres, we can achieve a more holistic understanding of eccentric HJs.

In Section \ref{sec: Methods}, we discuss the GCM and active magnetic drag prescription we use in this work. These models are cloudless, so we also introduce a 1D cloud model for comparison in Section \ref{subsec: 1Dcloud}.
In our results, we first compare a circular and eccentric model without magnetic effects to explore how eccentricity independently impacts winds and thermal structure (Section \ref{sec:circvseccen}). We then analyze how varying surface dipolar magnetic field strengths influence these atmospheric dynamics in Section \ref{subsec: mageffects}. We perform a comparison of relevant atmospheric timescales, such as the radiative and the advective timescales, in Section \ref{subsec: Timescales}. We also investigate the influence of clouds through the use of a 1D cloudy model in Section \ref{subsec: 1Dclouds2}. 
We additionally explore simulated emission spectra in Section \ref{sec:spectra} and study the effect of active magnetic drag on the net Doppler shifts of these spectra. We discuss our findings in Section \ref{sec:Discussion}. Finally, we summarize our conclusions in Section \ref{sec:Conclusion}.

\section{Methods} \label{sec: Methods}
\subsection{GCM} \label{subsec: GCM}

For this work, we utilize the 3D general circulation model called the RM-GCM, which solves the primitive equations of meteorology in the frame of the planet to simulate the physics of atmospheric circulation\footnote{\url{  https://github.com/emily-rauscher/RM-GCM}} \citep{2012RauscherGCM,Roman_2021,Malsky2021,Beltz2021}, coupled to the updated picket fence radiative transfer first coupled to this GCM in \cite{Malsky2024PF}. This picket fence approach \citep{ParmentierGuillotPFderive2014,Parmentier2015PF,Lee2022RT} considers opacities in five bands: two ``thermal'' bands which account for molecular and continuum opacity and three ``starlight'' bands which account for the incident stellar irradiation. The specific values for opacities in each of these bands are calculated from analytic coefficient tables published in \cite{Parmentier2015PF}, which take into account effective temperature, optical depth, metallicity, and incoming stellar flux. This picket fence approach has been shown to provide good agreement with more sophisticated radiative transfer schemes (such as the correlated-$k$ method) for Hot Jupiter atmospheres \citep{Lee2022RT}. 

The RM-GCM stands out amongst other hot Jupiter GCMs due to its inclusion of kinematic MHD through the use of an active drag timescale. This timescale, derived in \cite{Perna2010magdrag} from the non-ideal MHD equations, seeks to encapsulate the impact of the Lorentz force on the atmospheric dynamics. In the GCM, this drag is applied as a Rayleigh drag in the model's momentum equation in the form $\frac{-{\bf u}}{\tau_{mag}}$. To account for this loss of kinetic energy, a corresponding Ohmic dissipation of $\frac{u^{2}}{\tau_{mag}}$ is included in our thermodynamic energy equation, locally heating the atmosphere. First applied to HJs in \cite{RauscherMenou2013} and UHJs in \cite{Beltz2022a}, the timescale is as follows:
\begin{equation} \label{tdrag}
    \tau_{mag}(B,\rho,T, \phi) = \frac{4 \pi \rho \ \eta (\rho, T)}{B^{2} |\mathrm{sin}(\phi) | }
\end{equation}
where $B$ is the chosen surface/atmospheric magnetic field strength, $\phi$ is the latitude, and $\rho$ is the density. $\eta$ represents the magnetic resistivity,
\begin{equation} \label{resistivity}
    \eta = 230 \sqrt{T} / x_{e} \textnormal{ cm$^{2}$ s$^{-1}$ }
\end{equation}
where the ionization fraction $x_{e}$, here assumed to be much less than 1. This is calculated using the Saha equation \citep{1991Sato} 
\begin{equation} \label{Saha1}
    \begin{aligned} 
    x_{e} \equiv \dfrac{n_{e}}{n_{n}} & = \sum_{j} \dfrac{n_{e}}{n_{n}} x_{j}, \\ \dfrac{x_{j}^{2}}{1-x_{j}^{2}} & \simeq \dfrac{1}{n_{j}kT} (\dfrac{2\pi m_{e}}{h^{2}})^{3/2} (kT)^{5/2} \exp{(-\dfrac{I_j}{kT})} \\
    \end{aligned}
 \end{equation} 
where $n_{e}$, $n_{n}$, and $n$ are the number densities of electrons, neutrals, and total number density respectively.  $n_{j}$ represents the number density of each element $j$ with $I_{j}$ corresponding to the first ionization potential. $T$ is the temperature,  while $k$, $m_{e}$, and $h$ represent Boltzmann's constant, the mass of an electron, and Planck's constant. With this equation, we consider the first 28 elements of the periodic table, H to Ni, and add up the resulting values for a final ionization fraction. Other iterations of the kinematic MHD approach have used this form as well \citep{RauscherMenou2013,Beltz2022a,2024Wazny} For this approximation of resistivity, we assume that the electron number density and ion number density are approximately equal, which is valid for an atmosphere where the gas is essentially neutral \citep[see also][]{Laine2008, 2010PernaB, Menou_2012}.

We choose to model the magnetic field as a dipole aligned with the planet's axis of rotation. As a result, our active drag is applied solely in the east-west direction. This approach to magnetic drag allows our timescale to vary by over 10 orders of magnitude for a single pressure level in UHJs, largely due to the differences in thermal ionization, and thus resistivity, between the day and night sides due to their drastically different temperatures. Analysis of  \textit{Spitzer} phase curves indicate strong (often over 1000K) day-night temperature gradients for HJs and UHJs \citep{Bell2021,May2022}, highlighting the usefulness of our temperature dependent kinematic MHD approach.  

We finally note that our kinematic MHD approach is most appropriate for regions where the magnetic Reynolds number is less than 1, as we assume that any induced atmospheric magnetic field is smaller than the surface field strength \citep{Menou_2012,RauscherMenou2013}. The magnetic Reynolds number, $R_{mag} \approx \frac{U H}{\eta}$ \citep{Hindle2021observationalconsequences} is very small ($R_\mathrm{mag} \ll 10^{-5}$) for essentially the entire atmosphere, due to its moderate temperature. There are a few gridpoints in our simulations where $R_\mathrm{mag} \approx 1$, but this is confined to a small hot region in the deep ($>1~\mathrm{bar}$) atmosphere that has a minimal effect on the global atmospheric structure and post-processed spectra. Thus, our kinematic MHD assumptions are generally valid for this planet. 

Here, we present five models of an eccentric planet, TOI-150b, whose equilibrium temperature varies by over 400K throughout its orbit. We run each simulation with 65 vertical layers covering seven orders of magnitude in pressure from 100 to $10^{-5}$ bars at a resolution of T31 for 1000 orbits. Relevant planetary parameters are shown in Table \ref{tab:gcm_params}. We use values from the only two observational studies including this planet, \cite{Canas2019} and \cite{Kossakowski2019}. 

These papers disagree on mass and radius measurements, likely stemming from the fact that only \cite{Kossakowski2019} fit the eccentric orbit, while \cite{Canas2019} assumed a circular orbit--note also that the uncertainties for mass and radius between these papers do not overlap. However, both sets of planetary parameters keep the planet in the same mass regime. 
Here we choose to include parameters best suited for numerical stability from both papers, which led to the lower gravity, mass, and radius from \cite{Canas2019} and the larger semimajor axis from \cite{Kossakowski2019}. We do not expect different circulation patterns due to the different potential gravities of this planet, as gravity only vertically re-scales the dynamics with the the primitive equations used in this model \citep{Thomson2019}. This range of potential gravities will have slightly different photospheric pressures probed by our radiative transfer post-processing technique, but we expect this effect to be small. We show one circular model for comparison, and 4 eccentric (e=0.26) models with differing surface field strength (0G, 3G, 10G, and 30G). For the circular model, we assumed synchronous rotation while the eccentric models' rotation rates were chosen to be pseudo-synchronous, following \cite{Hut1981} and \cite{Kataria2013}. 


\begin{deluxetable*}{lcccc}
\caption{TOI-150b Chosen Model Parameters} 
\label{tab:gcm_params}
\tablehead{ \colhead{Parameter Name} & \colhead{Value} & \colhead{Units} & \colhead{Reference}}
\startdata
         Mass & 1.75 & M$_\mathrm{Jup}$ & \cite{Canas2019} \\ 
         Radius & 1.38 & R$_\mathrm{Jup}$  & \cite{Canas2019} \\
         Eccentricity & 0.26 & N/A  &  \cite{Kossakowski2019} \\
         Semimajor Axis & 0.070 & AU &  \cite{Kossakowski2019} \\
         Orbital Period & 5.86 & Days & \cite{Canas2019, Kossakowski2019} \\ 
         Pseudo-Synchronous Rotational Period & 4.14 & Days &  \cite{Hut1981} \\
\enddata
\end{deluxetable*}

\subsection{Post-Processing Radiative Transfer} \label{subsec: radtran}
In addition to the 3D GCMs presented, we post-process these models to generate simulated emission spectra. We follow the procedure described in \cite{Zhang2017} and \cite{Beltz2022b} to produce emission spectra throughout the planet's orbit. In short, the temperature and wind structure of the GCM is re-mapped to a constant altitude grid. At this point, we administer a ray-tracing radiative transfer scheme \citep[first described in][]{Zhang2017} utilizing the two-stream approximation for inhomogeneous multiple scattering atmospheres \citep{Toon1989} to calculate the emergent thermal spectra. For each line of sight column (one for each latitude and longitude value) we calculate the total optical depth by integrating over the local opacities for each wavelength point. 

We choose to generate both high-resolution ($R=100,000$) and low-resolution ($R=150$) spectra to explore the signatures of eccentricity and kinematic MHD in two resolution regimes. When calculating both sets of spectra, we assume local chemical equilibrium and solar abundances \citep{Lodders2003, Fastchem2018}. Our opacities are all taken from the ExoMol database \citep{exomol}.

For the high-resolution spectra, we choose the wavelength range 2.3-2.35 microns to conduct our post-processing on because CO is the dominant opacity source in that wavelength range. The CO abundance is expected to be nearly uniform around the planet, making it an excellent tracer molecule for dynamics \citep{Savel2023}. These spectra also contain opacities from H$_{2}$O, TiO, VO, K, and Na, though these species have a small contribution to the overall spectra compared to CO. We generate these spectra with and without Doppler broadening due to winds and rotation for our cross-correlation analysis.  

For our low-resolution spectra, we calculate spectra near apastron and periastron at a resolution of 5000, then convolve the resolution down to R$=150$. These spectra includes opacities from C$_{2}$H$_{2}$, CH$_{4}$, CO, CO$_{2}$, FeH, H$_{2}$O, H$_{2}$S, HCN, K, Na, NH$_{3}$, TiO, and VO. Given the low resolution, we do not generate spectra with Doppler effects, as winds cannot be constrained at low spectral resolution.

\subsection{1D cloud Modeling} \label{subsec: 1Dcloud}
In addition to the previously mentioned GCMs, we also explore the role of clouds and vertical mixing through a 1D cloud model. We model the planet using \texttt{EGP+} \citep{2021Mayorga}, which iteratively solves for an equilibrium radiative-convective solution by time-stepping an initial atmospheric structure. \texttt{EGP+} is based on the 1D radiative-convective equilibrium models of \cite{Marley1996, 2002Marley} and the time-stepping evolution of \cite{2014Robinson}. We assume solar metallicity and C/O ratio and do not cold trap refractory species, such as TiO and VO, which typically result in hot stratospheres \citep{Fortney2008} when temperatures are sufficiently high in the upper atmosphere. Planetary parameters are shown in Table \ref{tab:gcm_params}. We follow the same initialization and run time procedures as described in \cite{2021Mayorga} to ensure convergence and allow MnS and Cr clouds to condense where appropriate. We model three cloud structure cases, where we vary from large particles in vertically compact clouds to small particles in vertically extended clouds via the sedimentation efficiency, $f_\mathrm{sed}=3, 1, 0.5$, and the cloud free case. 

\section{Results} \label{sec: Results}
As this is the first 3D GCM to study the effects of magnetism on an eccentric planet, we will first isolate the effect of eccentricity alone by comparing our two drag-free models with and without a circular orbit. After isolating these differences, we then explore the effect of varying magnetic field strengths for an eccentric planet. 
\subsection{Comparison of Drag-Free Eccentric and Circular Models} \label{sec:circvseccen}
\begin{figure}
    \centering
    \includegraphics[width=4in]{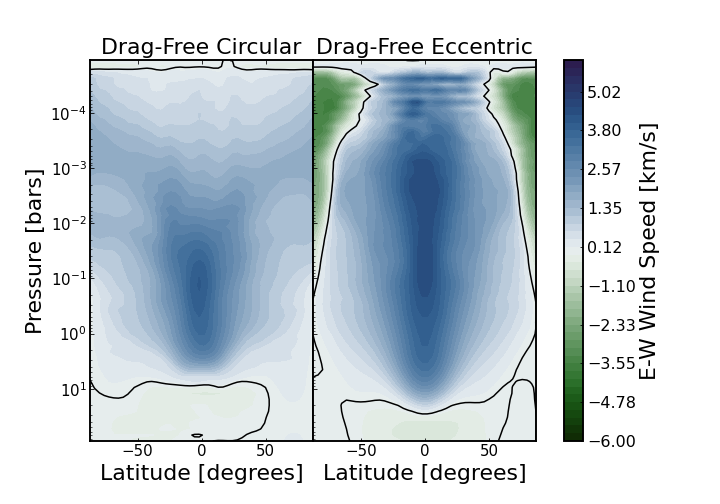}
    \caption{ The presence of an eccentric orbit results in the emergence of strong westward motion at high latitudes in the upper atmosphere in addition to the equatorial jet regularly found in circular models. Here we show zonal-mean zonal winds, plotted as a function of latitude over the modeled pressure region. When the orbit is non-circular, the equatorial superrotating jet  is narrower and stronger, particularly at pressures less than $\sim 0.01$ bar. We find that maximum wind speeds are generally faster in a non-circular orbit.} 
    \label{fig: TOI150bAVGPressureGrid}
\end{figure}

The inclusion of eccentricity results in significant changes to the atmospheric thermal structure and wind speeds. The variation in irradiation is the main driver for these changes, although the slightly-faster pseudo-synchronous rotation period additionally plays a role. We can immediately see one of the major consequences of the inclusion of eccentricity in Figure \ref{fig: TOI150bAVGPressureGrid}, where we show longitudinally averaged zonal wind speeds. The circular model shows the typical equatorial jet commonly seen in hot Jupiter models. For the eccentric model, we see the emergence of high latitude \textit{westward} motion in the upper atmosphere in addition to the expected equatorial jet. This westward motion was previously seen in \cite{Kataria2013}, but is stronger in the models presented here. This motion is a result of differences in the time-averaged structure of Rossby gyres between the two cases. These gyres are present in both models, but are smaller in size and faster in peak wind speed in the eccentric case. These smaller gyres have faster westward wind speeds near the poles, resulting in net westward flow at high latitudes. The eastward equatorial jet itself also reaches a higher peak speed and extends deeper into the atmosphere in the eccentric case. Additionally, this jet is narrower due to the decreased Rhines scale resulting from the faster rotation rate. 

\begin{figure*}
    \centering
    \includegraphics[width=5in]{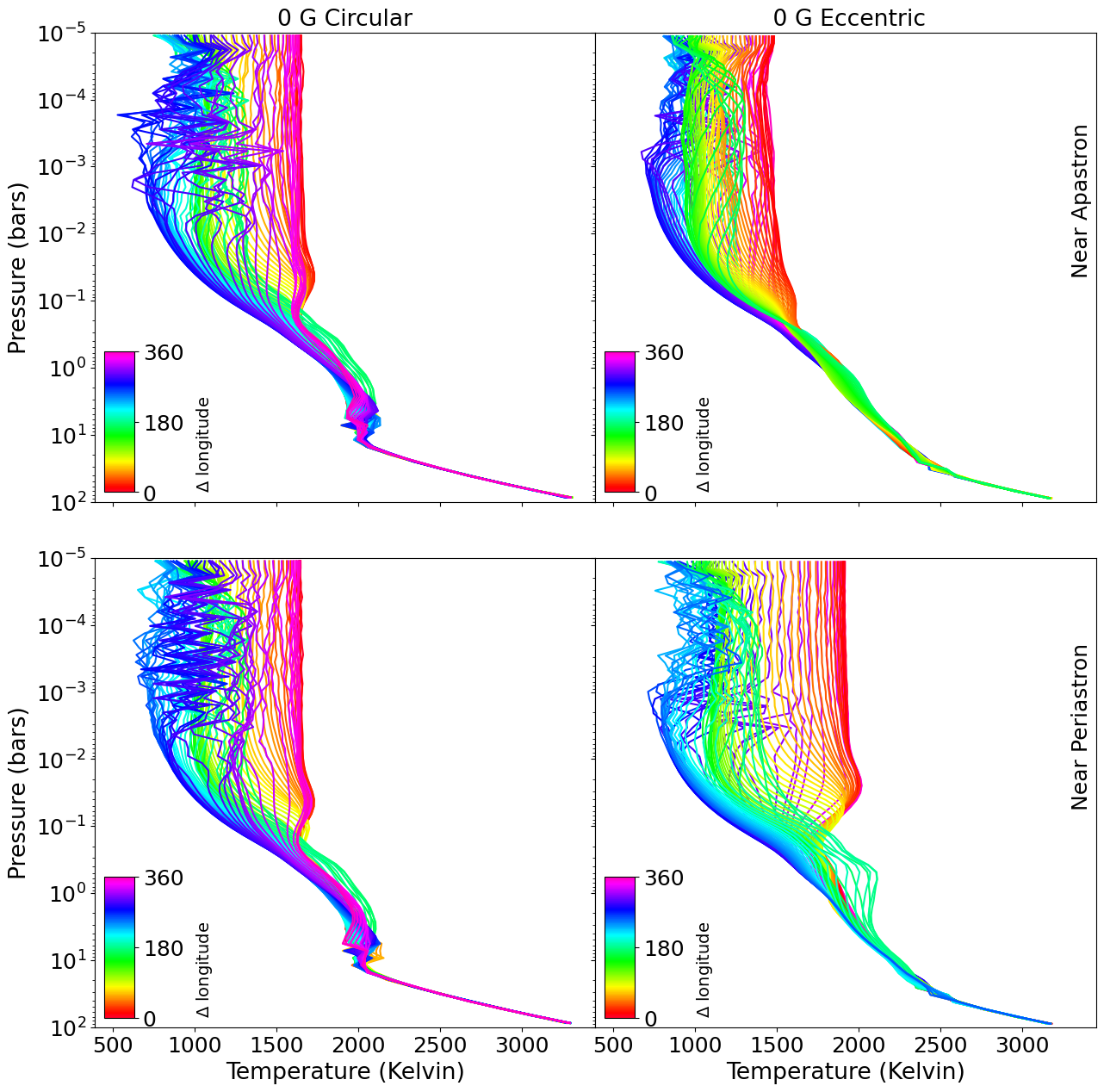}
    \caption{Eccentricity has a direct impact on the thermal profile of a hot Jupiter, which can result in thermal inversions that are only present for a portion of the orbit.  Here we present equatorial profiles of the 0G circular (left)  and eccentric cases (right) to show the impact an eccentric orbit has on the temperature-pressure profile of TOI-150b. The profiles are color-coded based on the distance from the substellar longitude (denoted as $\Delta$ longitude). Because of the variable orbital separation, the temperature-pressure profiles near apastron and near periastron differ greatly when eccentricity is applied. A thermal inversion occurs near periastron, where the planet receives the most stellar irradiation, but not near apastron, where the planet is much cooler. However, in the circular model, where e=0, there is a very slight inversion at the phases corresponding to near apastron and near periastron from the eccentric case. This is present throughout the entire orbit. }
    \label{fig: TOI150b0GTPCompare}
\end{figure*}

We next explore the differences in temperature structure as a result of eccentricity. Due to the time-variable irradiation received by the eccentric models over the course of its orbit, there are significant changes in the temperature structure over time as seen in Figure \ref{fig: TOI150b0GTPCompare}. Due to the variable irradiation, our eccentric model contains temperature inversions for only part of its orbit, particularly near periastron, where it receives the strongest irradiation. However, at cooler phases, no temperature inversions are present. The circular model, on the other hand, maintains a small temperature inversion for the entire orbit.

\begin{figure*}
    \centering
    \includegraphics[width=4.5in]{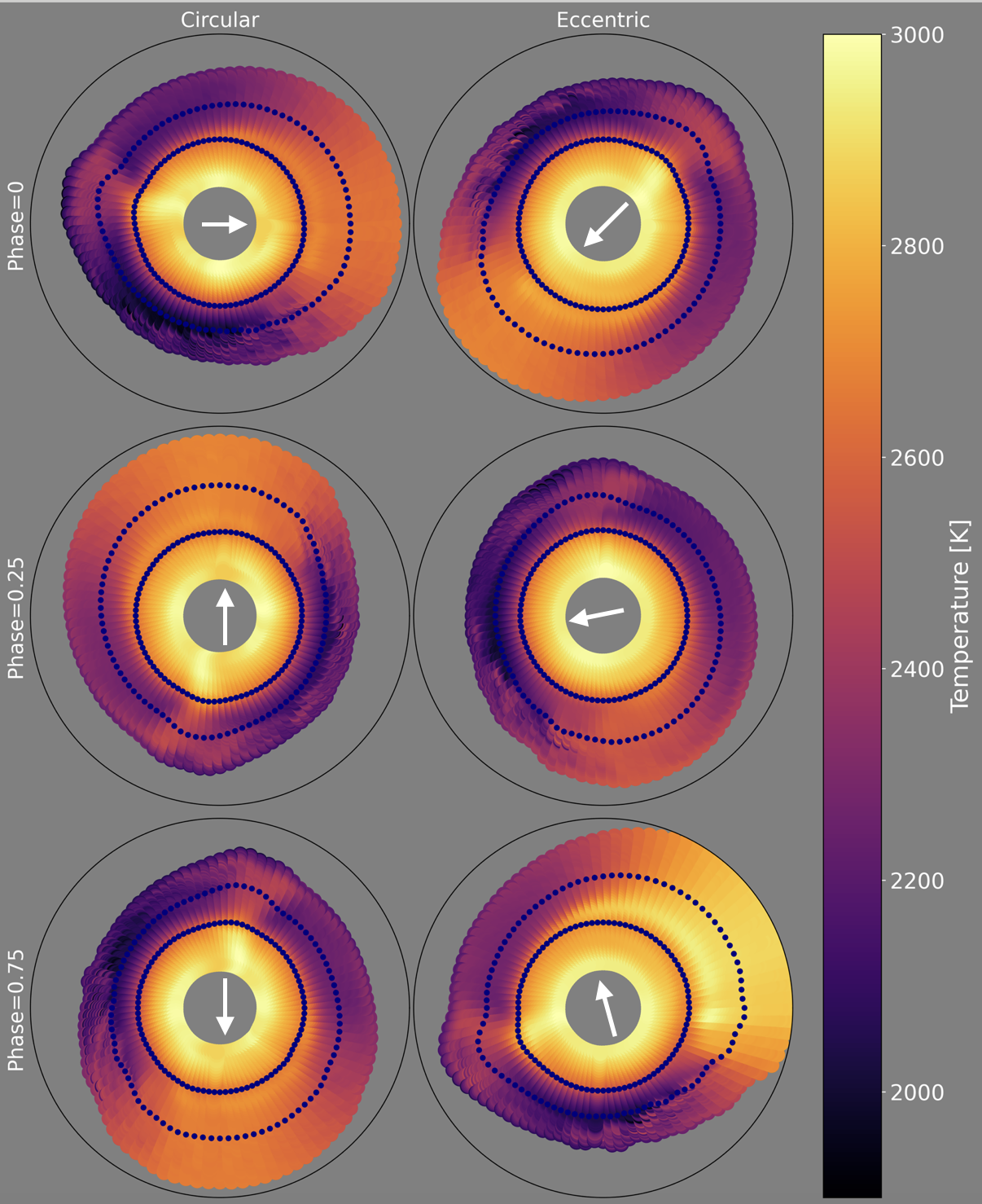}
    \caption{ At different phases of the planet's orbit, the atmosphere has varying day-night temperature structures. We present equatorial slices of TOI-150b's atmosphere at multiple phases (from top to bottom: transit, first quadrature, and third quadrature of the circular model  for both the circular (left) and eccentric (right) cases to showcase a top-down view of the planet's atmosphere. The white arrows at the center of each projection show the approximate direction of the host star at that phase. The temperature is projected on an altitude grid, with pressure levels of 0.1 bar and 1 millibar shown in the blue curves.  By plotting in altitude space, we can also see the effect of spatial variation in temperature affecting the scale height and physical extent of the atmosphere \citep{2022Savel}. For the circular model, the overall temperature structure is similar at each phase, just shifted due to the rotation of the planet. The eccentric model on the other hand shows stronger changes in temperature throughout its orbit. Additionally, the eccentric model has a faster rotation rate then the circular model, which further alters the circulation structure.  
    } 
    \label{fig: TOI150beqslices}
\end{figure*}

Another way to visualize the differences between the circular model and the eccentric model can be seen in the equatorial slices shown in Figure \ref{fig: TOI150beqslices}. The circular model has similar temperature structures at each phase shown, just rotated for the different phases. The eccentric model's temperature structure is significantly different at each phase due to the planet's rotation, resulting in large variations in the apparent size due to the global temperatures increasing from apastron to periastron. 
In summary, the inclusion of eccentricity in our drag free models brings about high latitude westward motion and a temperature structure that varies significantly over a single orbit, showing temperature inversions during the phases where the planet is the hottest and a completely non-inverted atmosphere during regions of the orbit where the planet receives the lowest stellar irradiation. These results are similar to that of previous works, such as \cite{Kataria2013} and \cite{2014Lewis}.

\subsection{Magnetic Effects} \label{subsec: mageffects}
\begin{figure*}
    \centering
    \includegraphics[width=6in]{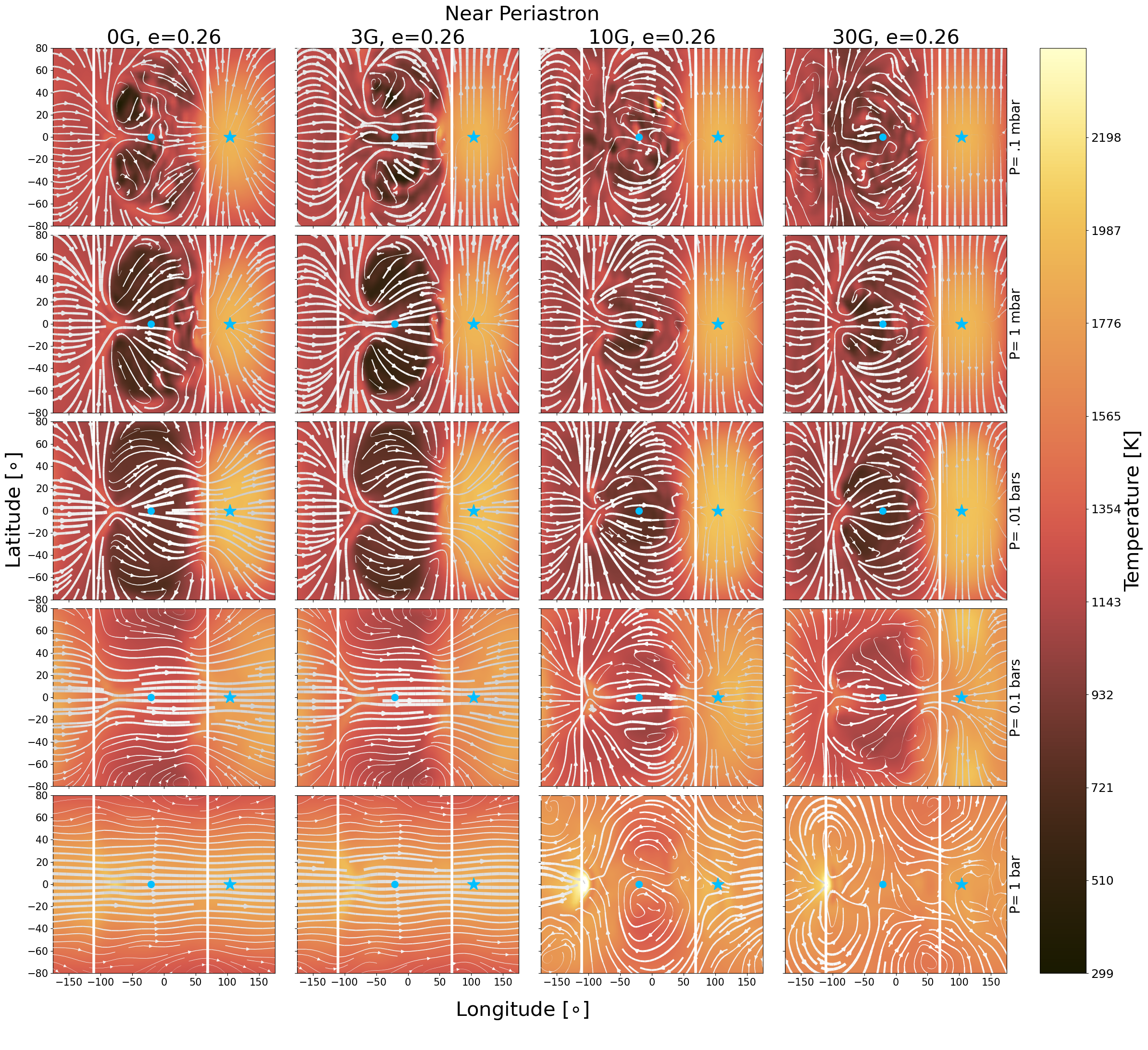}
    \caption{ The inclusion of active magnetic drag alters the temperature and wind structure, particularly in the upper atmosphere. Here we show maps of temperature and wind streamlines of TOI-150b with different levels of dipolar magnetic field strengths (columns) at a phase of 300$^{\circ}$, near periastron, where there is a thermal inversion present. The Earth-facing hemisphere is between the two white solid lines, and the substellar point is denoted by the star. The subobserver point is denoted by the circle. 
    Note that non-zero eccentricities result in hotspots that are no longer centered on the substellar point due to the lack of synchronous rotation. The presence of magnetism reduces the strength of the zonal winds, particularly on the day side, leaving primarily meridional winds present in the upper atmosphere.  } 
    \label{fig: StreamlineAll4NoDiff75}
\end{figure*}

We now explore how the addition of active magnetic drag shapes the atmosphere of an eccentric hot Jupiter.  
We show the temperature structure for all our magnetic models near periastron in Figure
 \ref{fig: StreamlineAll4NoDiff75}, and the corresponding structure near apastron in Figure \ref{fig: StreamlineAll4NoDiff26} in the appendix. The influence of our active drag is most readily seen in the hottest part of the upper atmosphere. The regions near the substellar point in the active drag cases show evidence of magnetic circulation, where the  winds follow magnetic field lines \citep{Batygin2013}. That is, they show strong winds in the north-south direction, resulting in poleward flow in the hottest regions on the dayside. We find that the strength of the initial magnetic field determines the extent to which this flow persists in the deeper atmosphere. Near apastron (see Figure \ref{fig: StreamlineAll4NoDiff26}), the choice of dipole field strength is less influential, as the cooler temperatures result in weaker magnetic drag. At both phases presented, the 10G and 30G cases show a different flow pattern at pressures greater than 0.1 bar, due to the weakened jet structure discussed below. 

 A small temperature inversion is present near the substellar point during the full orbit when the planet is on a circular orbit. The introduction of eccentricity results in a temperature inversion that is present for only a fraction of the orbit (see Figure \ref{fig: TOI150b0GTPCompare}). Each eccentric model exhibits temperature inversions, however for differing fractions of the orbit. In all models, this temperature inversion remains the strongest near the substellar point. Near periastron, longitudes as large as $30^\circ$ away from the substellar point still show smaller, but significant inversions. The 0G model retains its radiatively driven temperature inversion  for a longer time after periastron  than any of the active-drag models. This inversion also has the latest onset of any models explored. This can be understood by comparing the advective and radiative timescales in Figure \ref{fig: timescales}. The inclusion of active drag results in a longer advective timescale and thus a dominant radiative timescale. This results in the thermal structure predominantly responding to irradiation rather than heat transport.   In addition, the inclusion of active drag results in a slightly stronger inversion compared to the drag-free case, as seen in Figure \ref{fig: TPProfNearPeriAll4Form2}. 


\begin{figure*}
    \centering
    \includegraphics[width=8in]{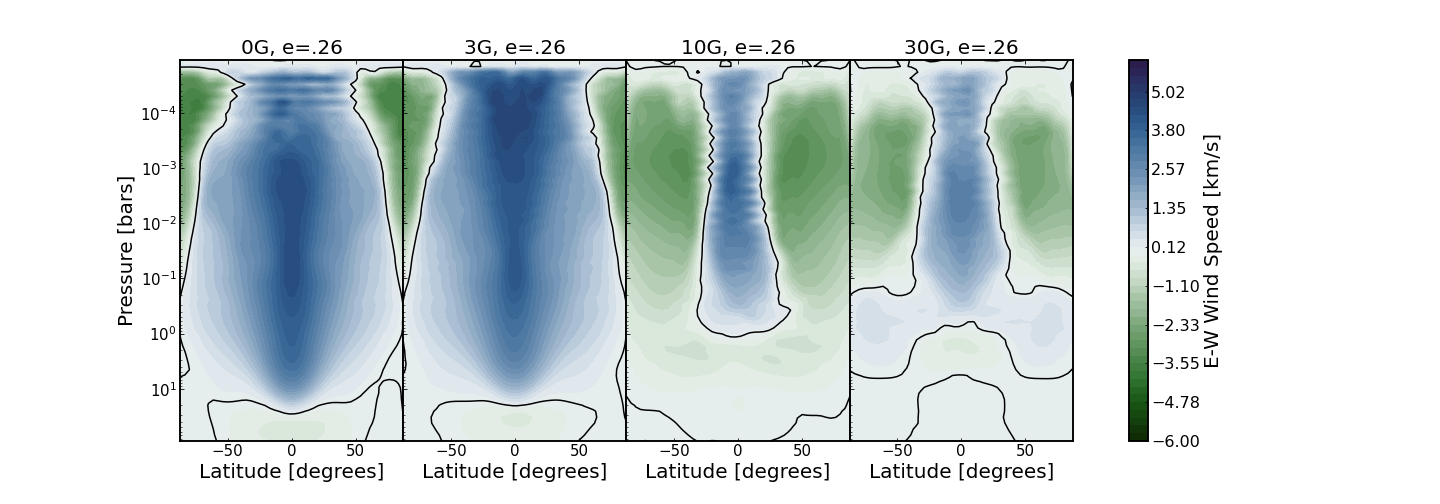}
    \caption{ The inclusion of kinematic MHD can dramatically alter the strength and extent of the equatorial jet. Here we show longitudinally averaged zonal winds plotted as a function of latitude and pressure. The zonal winds are averaged through each output phase over the course of one orbital period. The equatorial jet becomes weaker and narrower as the surface field strength increases. For these stronger magnetic field cases, we also find that the high latitude westward motion previously discussed extends deeper in the atmosphere and persists closer to the equator.  } 
    \label{fig: TOI150bAVGPressureGridall4}
\end{figure*}

We also explore the effect of varying dipole field strength on the equatorial jet, as shown in Figure \ref{fig: TOI150bAVGPressureGridall4}. At all field strengths applied, we still see the presence of an equatorial eastward jet in the zonal-mean winds, but this jet is much weaker for the 10G and 30G cases. Interestingly, the jet in the 3G case is stronger in the upper atmosphere than the 0G case. The high latitude westward motion identified in Section \ref{sec:circvseccen} is present in all of our eccentric models. For the 10G and 30G cases, this westward motion extends to the mid-latitudes, while it is strictly confined to the uppermost latitudes in the 0G and 3G cases. 

\begin{figure*}
    \centering
    \includegraphics[width=6in]{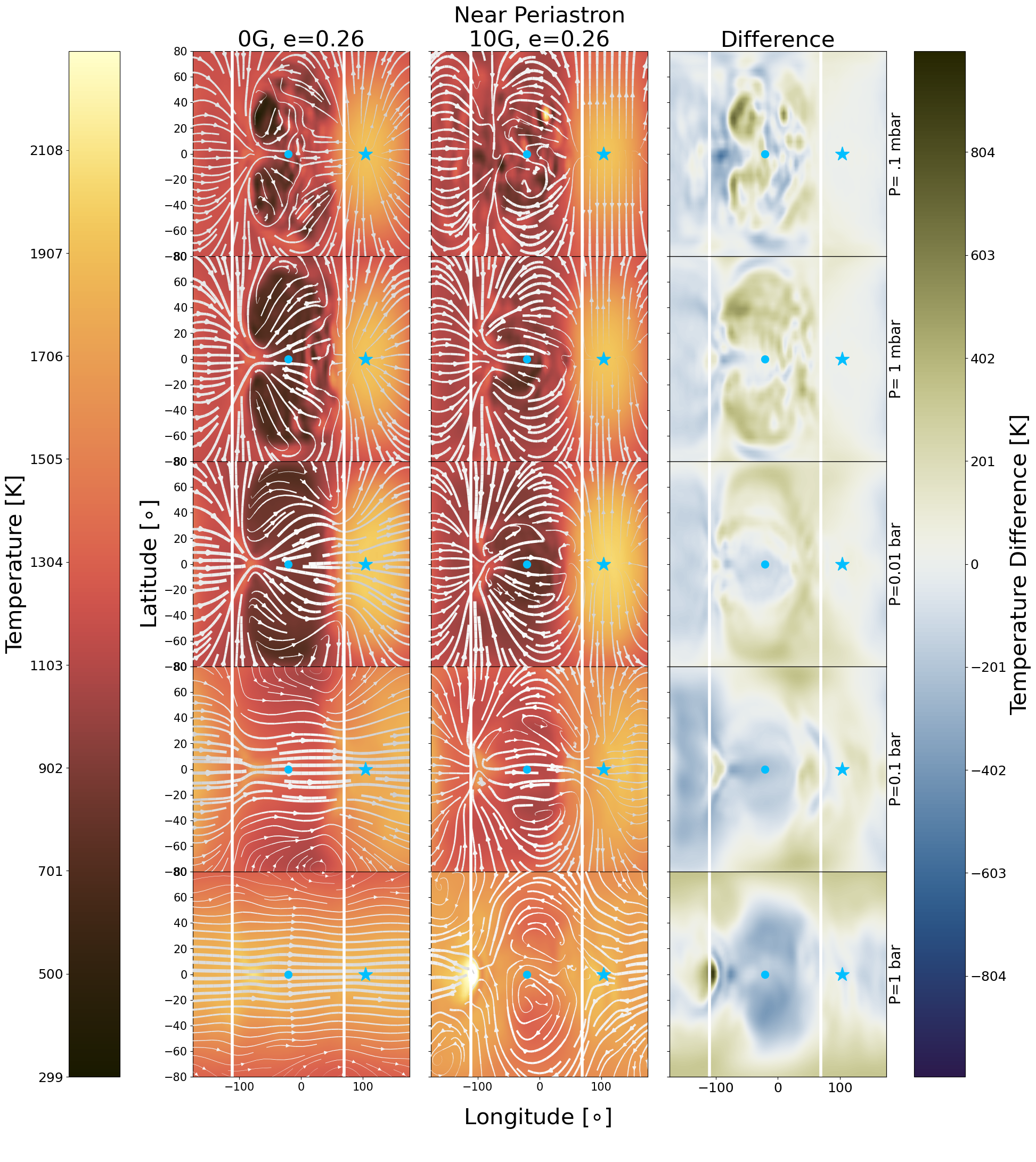}
    \caption{ Increasing magnetic field strength will alter the thermal structure of an eccentric hot Jupiter. Here we show the difference between the 10G and 0G eccentric cases at a phase near periastron (right hand column) alongside their near periastron temperature and wind maps. The Earth-facing hemisphere is between the two white solid lines, and the substellar point is denoted by the star. The subobserver point is denoted by the blue circle.
    At lower pressures (higher in the atmosphere), the 10G case is warmer west of the hotspot and cooler to the east, due to the weaker Rossby gyres. } 
    \label{fig: TOI150b10Gvs0GNearPeri}
\end{figure*}

We can see the difference in temperature structure between our drag free and active drag cases more clearly in Figure \ref{fig: TOI150b10Gvs0GNearPeri} (with the corresponding maps at apastron shown in Figure \ref{fig: TOI150b10Gvs0GNearApas} in Section \ref{sec:Appendix}). In the upper atmospheres, both models reach roughly the same temperature at the hotspot. The 10G model is generally warmer than the 0G case west of this point and slightly cooler east of the hotspot. This is due to the low temperature pair of Rossby gyres occupying the mid-latitudes in the 0G case. A local thermal minimum exists near the equator just west of the hotspot in each of the eccentric models. In the active drag cases, the jet is slowed from the applied drag, increasing the local divergent flow and resulting in adiabatic cooling near the equator due to upwelling. 

\begin{figure}
    \centering
    \includegraphics[width=3.5in]{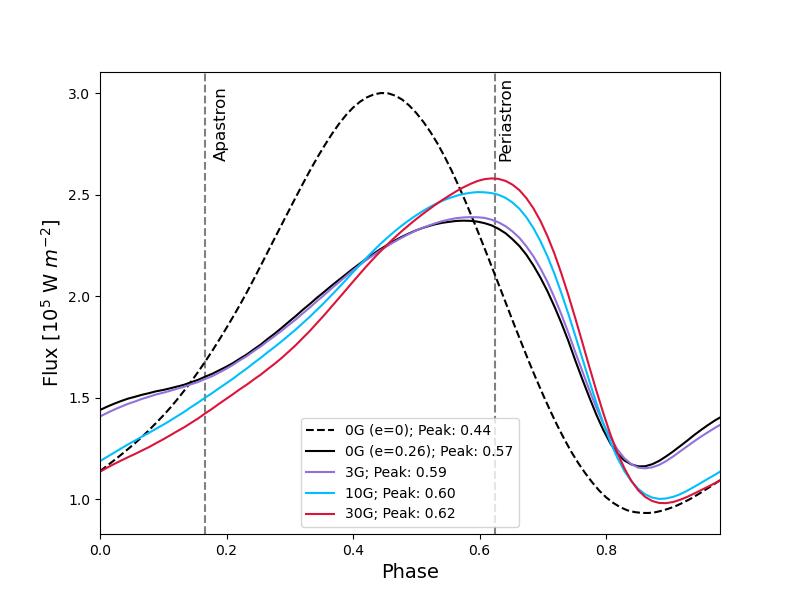}
    \caption{ The inclusion of eccentricity and magnetic drag influences the shape and peak of corresponding phase curves. All the eccentric models reach peak flux prior to periastron. The higher the magnetic field strength modeled, the closer the peak flux is to periastron. These phase curves were calculated from the model's outgoing bolometric thermal emission. Since the eccentric models complete more than one rotation over a single orbit, the fluxes at the beginning and end of the phase curves may be offset. } 
    \label{fig: phasecurves}
\end{figure}

The difference in temperature structures will also manifest in phase curves, as shown in Figure \ref{fig: phasecurves}. The circular model shows the typical hot Jupiter behavior of reaching a peak flux slightly before secondary eclipse due to the presence of the superrotating equatorial jet. The effect of eccentricity is immediately seen in the shape and location of peak flux of the curves. Our eccentric models all reach a peak flux slightly before periastron and the inclusion of magnetic drag results in the peak amplitude occurring later in phase. The strongest field tested, 30G, reaches its peak closest to periastron.

\subsection{ Comparisons of Relevant Timescales} \label{subsec: Timescales}

    \label{tab:timescales_table}
\begin{figure}
    \centering
    \includegraphics[width=3.5in]{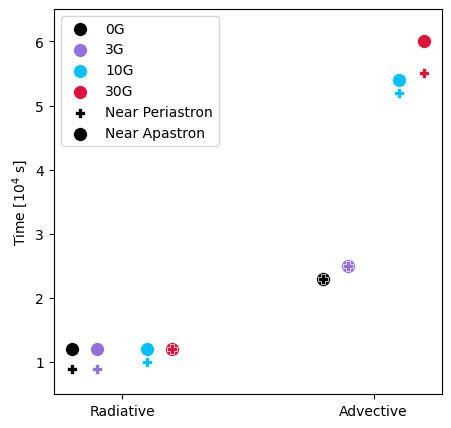}
    \caption{ Radiative and advective timescales for our eccentric models using Equations \ref{radtimescale} and \ref{advtimescale} respectively. Periastron values are shown with crosses and apastron values are shown with circles. Each timescale is calculated at a pressure of 150mbar, the approximate location of the IR photosphere. For representative values of temperature and wind speed, we averaged over the 9 points closest to the substellar longitude.} 
    \label{fig: timescales}
\end{figure}
To further understand how atmospheric conditions vary throughout the eccentric orbit and better gauge the impact of magnetic effects, we estimate the radiative and advective timescales near periastron and near apastron for each model in Figure \ref{fig: timescales}. For both timescales,   we average the temperature and east-west wind speed over the model region $\pm 10^{\circ}$ from the substellar point at a pressure level roughly corresponding to the IR photosphere.\footnote{Note that this method is an estimate of the radiative timsecale. For a more robust calculation method, see \citep{Showman2008}} The radiative timescale, which represents the time required for the stellar irradiation to be re-emitted by the planet, was estimated as follows:

\begin{equation} \label{radtimescale}
    \tau_{rad} = \frac{P}{g}\frac{c_{p}}{4\sigma T^{3}}
\end{equation}

where $P$ is pressure, $g$ is surface gravity, $c_p$ is specific heat capacity $\sigma$ is the Stefan-Boltzmann constant, and $T$ is the temperature \citep{2010Showman}.

As expected, the radiative timescale for each model is shorter at periastron due to the increased temperatures. Each model reaches similar maximum temperatures, resulting in a small variance between the periastron radiative timescales. As the field strength is increased, the difference between the apastron and periastron time scale becomes smaller.   

We additionally calculate the advective timescale, which should have a stronger variance between our drag-free and magnetically active models due to the dependence on wind speed. This timescale represents amount of time it takes for winds to advect heat in the atmosphere and is calculated as follows:

\begin{equation} \label{advtimescale}
    \tau_{adv} = \frac{R}{U}
\end{equation}

where $R$ is the radius of the planet and $U$ is the zonal wind speed \citep{2010Showman}. Looking again at Figure \ref{fig: timescales}, we note that the 10G and 30G models have significantly longer timescales than the 3G and drag-free model. This follows expectations, as the equatorial jet for the 10G and 30G models are significantly slowed by the inclusion of our active magnetic drag (see Figure \ref{fig: TOI150bAVGPressureGridall4}). These two models with the weaker jet also show a greater difference in the advective timescale between apastron and periastron. This suggests that the strength of the disrupted ``jet'' can vary over the course of a single orbit. 

\subsection{1D Cloudy Models} \label{subsec: 1Dclouds2}
\begin{figure}
    \centering
    \includegraphics[width=3.5in]{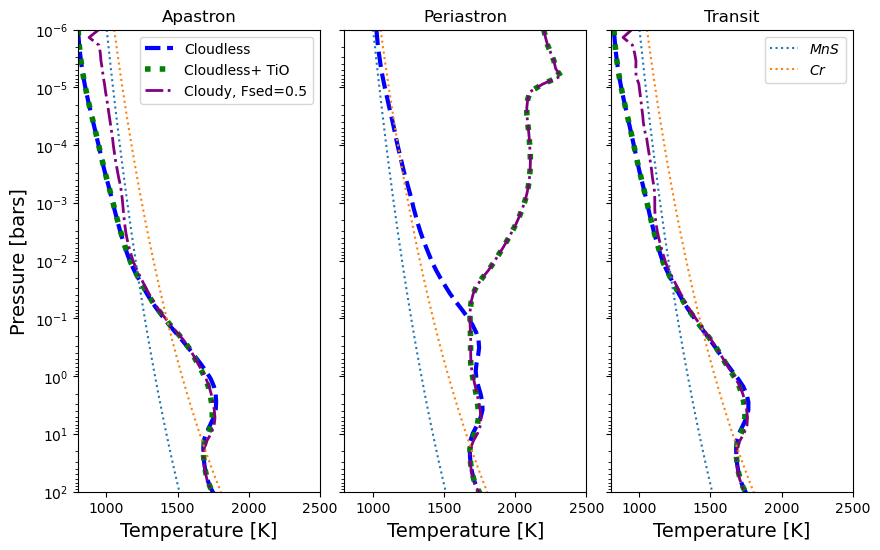}
    \caption{ Clouds result in slight upper atmospheric warming at transit and apastron due to the increase in opacity that leads to a greater deposition of energy, but have a minimal effect on the temperature structure at periastron. The inclusion of TiO creates temperature inversions near periastron and has more of an impact on the temperature structure for this planet than the inclusion of clouds. Cloud condensation curves are shown as dotted lines. More cloud species form at apastron and transit, when the model is cooler at pressures $\lesssim$ 0.01 bar.}
     
    \label{fig: 1D tprofs}
\end{figure}

The 3D GCMs shown are cloudless models. In order to gauge the potential effects of clouds, we ran a series of independent 1D models with and without clouds to explore the temperature structure and resulting spectra. We also included a cloudless model with TiO to gauge the potential influence of this absorber in the upper atmosphere. We present the T-P profiles at apastron, periastron, and transit in Figure \ref{fig: 1D tprofs}. At apastron and transit, all cases are similar except for a slightly hotter upper atmosphere in the cloudy models due to the increased opacity in the upper atmosphere. During this period of the planet's orbit, the presence of TiO doesn't affect the temperature structure and all profiles are non-inverted. We expect clouds of MnS, and Cr to form at pressures $\lesssim 0.01$ bar near apastron. Near periastron, we find that the inclusion of TiO induces temperature inversions. Because of the globally higher temperatures during periastron, clouds are only possible in the deep atmosphere.

Overall, we find from our 1D models that cloud coverage should be variable throughout the orbit of TOI-150b and they may disappear altogether at periastron. When the planet is near apastron, clouds can slightly warm the upper atmosphere. The inclusion of TiO as an opacity source has a greater impact on the temperature structure  than the inclusion of clouds. 

\subsection{Emission Spectra} 
\label{sec:spectra}
To understand the impact of eccentricity and our active magnetic drag on observables, we post-process our models as described in Section \ref{subsec: radtran} at two different resolutions. 
\subsubsection{High-Resolution Spectra}
\label{subsec: simspec}
\begin{figure} 
    \centering
    \includegraphics[width=3.5in]{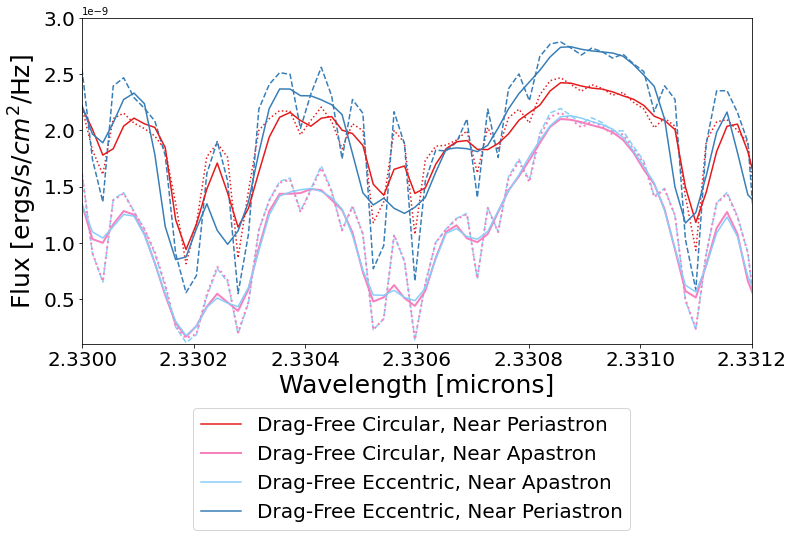}
    \caption{ The changing stellar flux associated with an eccentric orbit causes stronger atmospheric variation throughout the orbit than the circular model, resulting in a wider range of emission spectral fluxes. In this plot, we show spectra from the eccentric and non-eccentric drag-free models near apastron and periastron. Doppler broadened spectra---including effects from winds and rotation---are shown in the solid curves, with the dashed lines showing the spectra without Doppler broadening. The eccentric model has a larger difference in continuum flux between  periastron and apastron than the circular model due to the changing distance from the host star. }
    \label{fig: 0Gcomp}
\end{figure}

We generated high-resolution emission spectra between 2.3-2.35 microns for 24 phases in each model's orbit to explore the impact of magnetic drag on spectral features and Doppler broadening. We show spectra from the circular and eccentric hydrodynamic models in Figure \ref{fig: 0Gcomp}.

Near periastron, the eccentric model has a higher continuum flux, due to the increased stellar irradiation. 
Differences between the spectra at apastron and periastron for the circular model are due to the tidally locked nature of the planet, but are smaller than that of the eccentric model. The two apastron spectra have nearly the same continuum flux, meaning their hemispherically-averaged photospheric temperatures are similar. The wind structures vary between these two cases, as shown by the differences in the Doppler broadened spectra (see Figure \ref{fig: 0Gcomp}).

\begin{figure}
    \centering
    \includegraphics[width=3.5in]{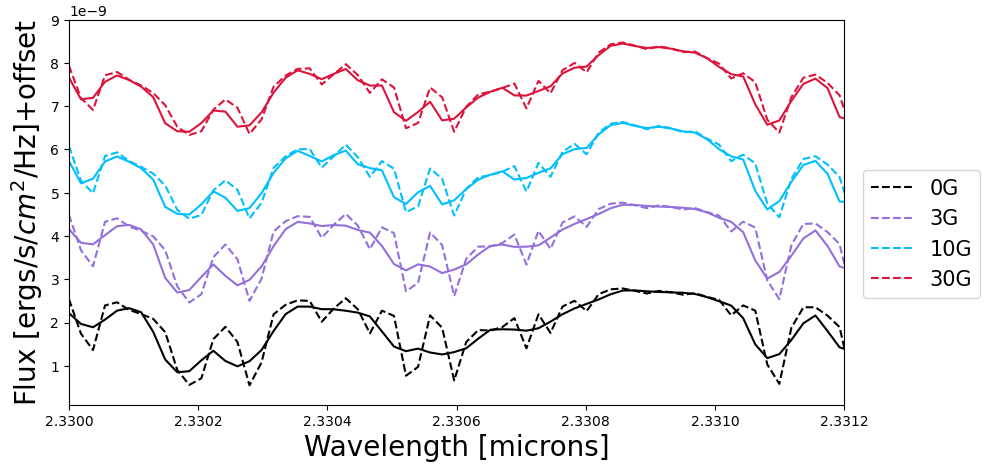}
    \caption{ Changing the magnetic field strength results in differing levels of broadening in the resulting spectra. We display each of the post-processed eccentric simulations with a vertical offset applied to distinctly see the characteristics of each emission spectra. We show the spectra near periastron as magnetic effects are expected to be the dominating force behind atmospheric structure at this phase. Spectra including Doppler effects (due to winds and rotation) are shown in solid lines while spectra without Doppler effects are shown in dashed curves. 
    There is a reduction in the broadening effect seen in the Doppler shift at high levels of magnetism as is demonstrated in Figure \ref{fig: dopbroad}. Compared to the 0G and 3G cases, the 10G and 30G cases show significantly less Doppler broadening. This is because the wind strength is substantially reduced at higher field strengths, due to the increased drag.}
    \label{fig: OffsetAllModels}
\end{figure}

\begin{figure} 
    \centering
    \includegraphics[width=3.25in]{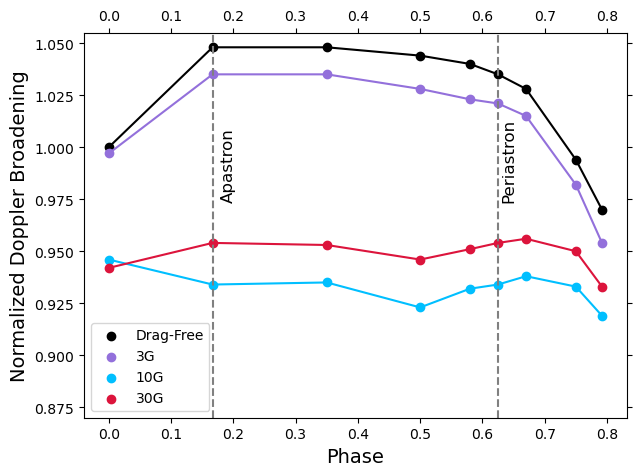}
    \caption{ The extent of Doppler broadening changes when different magnetic field strengths are applied. Here we present the Full-Width 80\%-Max of the cross-correlation curve normalized to the drag-free case at transit for each of our eccentric models for a variety of phases. Overall, the most strongly dragged models have less broadening at all phases compared to the drag-free and 3G model. Additionally, the 0G and 3G model show a larger range in broadening thoughout orbit compared to the strongly dragged models.  } 
    \label{fig: dopbroad}
\end{figure}

We explore how emission spectra varies with our chosen field strength in Figure \ref{fig: OffsetAllModels}. We plot the Doppler-broadened spectra in solid lines on top of the spectra without Doppler effects shown in dashed lines. Visually,  there appears to be less Doppler broadening as the magnetic field strength increases. To explore this further, we cross-correlated each Doppler on spectra with its corresponding Doppler off spectra. We estimated the amount of Doppler broadening by calculating the full width at 80\% maximum of the cross-correlation curve. We then normalized these values to the drag free eccentric model at an orbital phase of 0$^{\circ}$. These values are presented in Figure\ref{fig: dopbroad}. 
The inclusion of any amount of drag results in a decrease in the line broadening. Notably, the 10G model has the least amount of broadening at each phase, showing even less broadening than the 30G model except for at phase=0, where the two models have very similar levels of broadening. 
Furthermore, the level of broadening in the eccentric models near periastron is consistently smaller than those near apastron. This can also be explained by the shift in viewing angle and the fact that higher temperatures result in an increase in magnetic drag effects, leading to a decrease in Doppler broadening.  

\begin{figure*}
    \centering
    \includegraphics[width=6in]{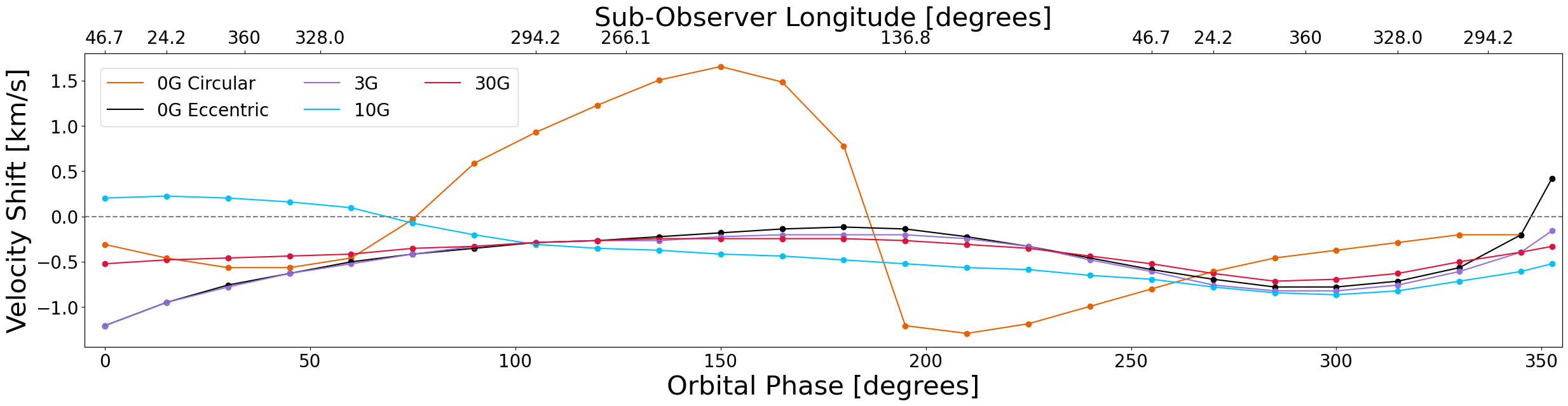}
    \caption{ Magnetic field strength and eccentricity have substantial impacts on the wind speeds and corresponding Doppler shifts of the planet. We cross-correlate our high-resolution Doppler on spectra with our high-resolution Doppler off spectra and show the peaks of the cross-correlations as a function of phase for each of our models. For circular orbits, one would expect the Doppler shifts presented at the beginning and end of orbit to be nearly identical. However, the eccentric cases have pseudo-synchronous rotation, thus we are not looking at the same hemisphere of the planet at the beginning and end of the orbit. A better comparison is to look at phases with the same subobserver longitude, which is denoted on the upper x-axis. For example, at phase 0 degrees and at phase 255 degrees in the planet's orbit, the observer is looking at the same hemisphere. 
    We note that the 10G model shows the largest difference in net Doppler shifts when looking at the same hemisphere over a single orbit, implying variation in the temperature and wind structure of the planet.}   
    \label{fig: AllModelsCC}
\end{figure*}

To understand how the disc-integrated wind pattern changes as function of phase, we calculate the net Doppler shift as a function of phase, shown in Figure \ref{fig: AllModelsCC}. To do so, we cross-correlated each Doppler broadened spectra with its corresponding spectra without Doppler broadening and found the shift associated with the peak of the cross-correlation function which  corresponds to the spectra's net Doppler shift. 
The drag-free circular case is both substantially redshifted and blueshifted at various points along its orbit. For a majority of the orbit, the eccentric models display a slight net redshift for this wavelength range. Exceptions to this are the 10 G case being mildly blueshifted in the beginning of its orbit and the drag-free eccentric case showing a slight blue-shift at the end of its orbit.
We find that the 10G and 30G cases have the lowest magnitude velocity shifts. This aligns with our expectations, as higher magnetic field strengths reduce the strength of wind features, resulting in smaller Doppler shifts.
 
For the circular case, since the orbit is assumed to be spin-synchronized, the velocity shift is nearly identical at the beginning and end of the orbit, as the same hemisphere is in view. The pseudo-synchronous nature of the eccentric models means the planets will complete a full rotation prior to the end of the orbit, 
resulting in the observer viewing different parts of the planet at the beginning and the end of one orbital period, meaning that the net Doppler shift found at the beginning and end of orbit won't necessarily match. However, this also means that some longitudes will be seen twice by the observer during one orbit. We include the sub-observer longitude on the top axis of Figure \ref{fig: AllModelsCC}. By examining the same sub-observer longitude at two different orbital phases, we note that the net Doppler shifts vary the most for the 10G model, while the other cases differ by less than 0.5 km/s when the same hemisphere is facing Earth. Current observational methods of constraining phase-resolved Doppler shifts are likely unable to reach this level of precision.\footnote{Recent phased-resolved Doppler shifts in emission from \cite{Pino2022} reach a precision of a few km/s. The errors can be smaller in phased-resolved transmission data, see \cite{Borsa2021} }  Between all the models, the 10 Gauss case has the largest deviation in net velocity shift---indicative of atmospheric variability over the course of a single orbit. 

\begin{figure*}
    \centering
    \includegraphics[width=6in]{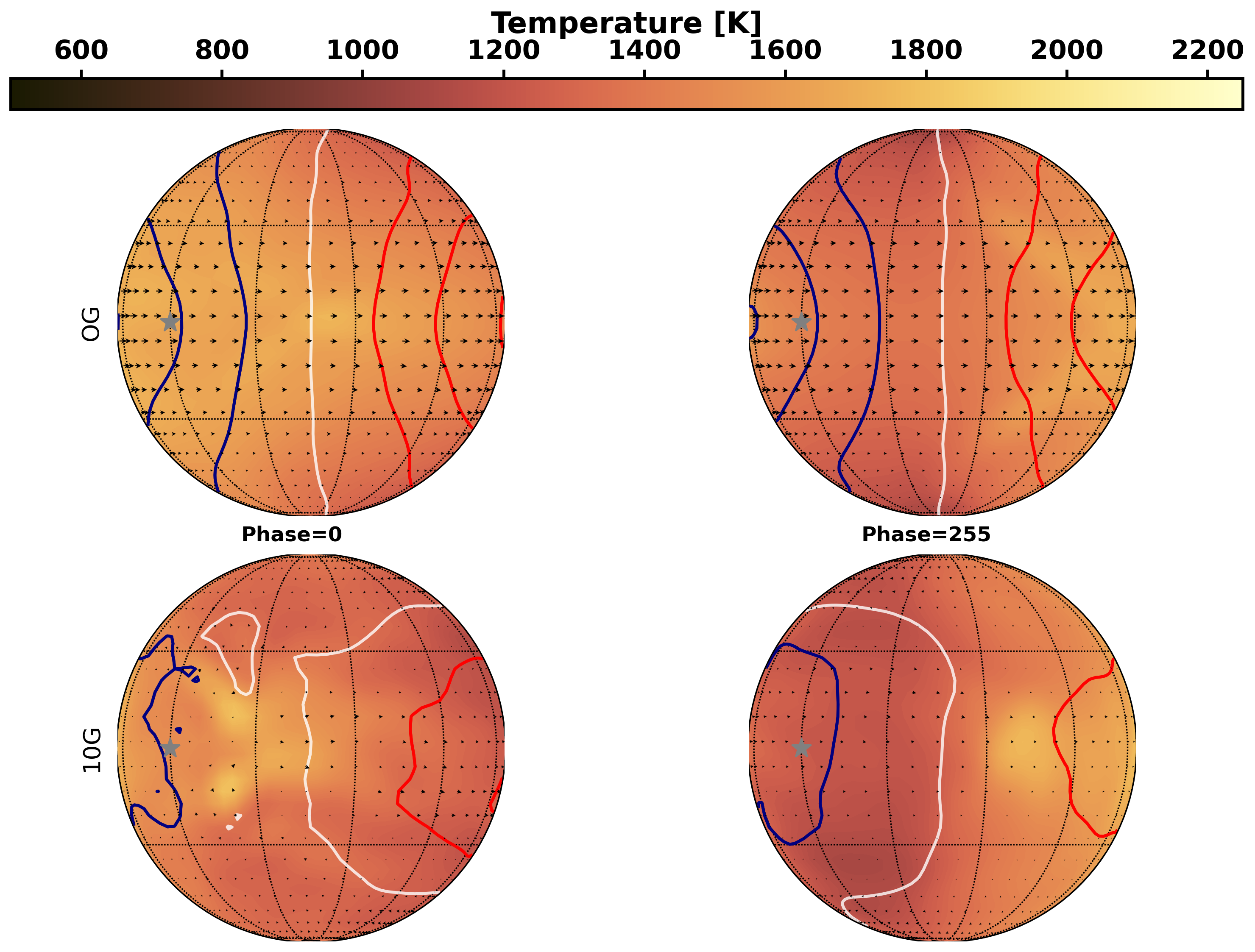}
    \caption{ Due to variability within a single orbit, the same hemisphere of the planet can show significantly different structures at two different phases. Here, we show orthographic projections near the photosphere for the 0G and 10G eccentric models at two phases where the same hemisphere is visible. The blue and red contours show lines of constant blueshifting/redshifting in increments of 2 km/s and the white contour shows where the line of sight velocity is 0. The grey star indicates the location of 0 longitude. Looking at the 10G case, we can see stronger variation in the wind structure at these two phases than the 0G model, due to the inclusion of our locally calculated magnetic drag.     } 
    \label{fig: orthoprojections}
\end{figure*}

 We can visualize this variability in Figure \ref{fig: orthoprojections}, which shows photospheric orthographic projections of our 0G and 10G eccentric model at  two phases with the same subobserver longitude. The 10G case shows significant variations in the line of sight velocity contours between these two phases while the contours for the 0G case look largely similar. This difference in wind structure helps drive the differences in net Doppler shifts presented in Figure \ref{fig: AllModelsCC}.

\subsubsection{Low-Resolution Spectra}

\begin{figure}
    \centering
    \includegraphics[width=3.5in]{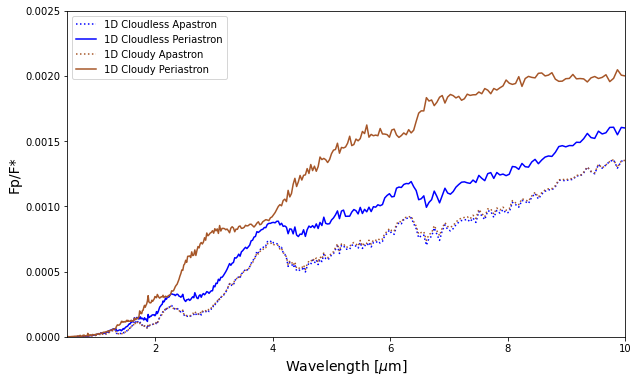}
    \caption{ The presence of clouds in a one-dimensional model results in a significantly increased flux at periastron, but have minimal effects at apastron. We show a comparison of the 1D Cloudy and cloudless emission spectra at a resolution of $R=150$. The inclusion of clouds causes a warming affect in the upper atmosphere at periastron (see Figure \ref{fig: 1D tprofs}) resulting in a higher flux ratio at all wavelengths. 
    Near apastron however, the impact of clouds is minimal. The two spectra never differ by more than $5\%$, due to their extremely similar temperature profiles at this phase.}
    \label{fig: 1D CloudyCloudless}
\end{figure}

We additionally generated low-resolution (R$\approx$150) emission spectra from the 1D models and a subset of our kinematic MHD models. We first compare the 1D cloudy and cloudless models in Figure \ref{fig: 1D CloudyCloudless}, to isolate the impact of clouds. At periastron, where the inclusion of clouds resulted in a hotter upper atmosphere, the cloudy model shows higher flux at all wavelengths. At apastron, the spectra are nearly identical due to the similar temperature profiles. Small variations between the spectra exist shortward of 2 microns, but the flux ratios do not differ by more than $5\%$. The influence of clouds on the emission spectra of the planet is thus phase-dependent, with the strongest influence occurring at periastron. 

\begin{figure}
    \centering
    \includegraphics[width=3.5in]{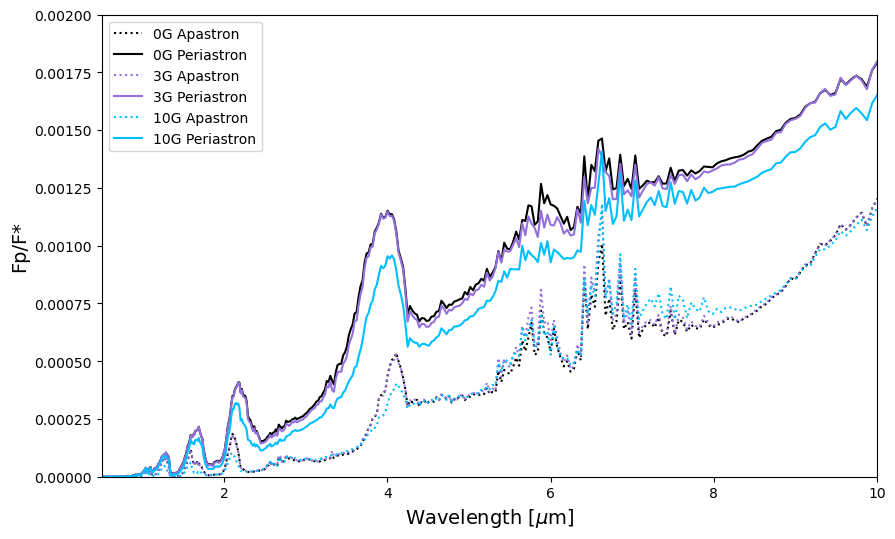}
    \caption{ The inclusion of magnetic fields, as well as 3D effects, can impact the strength and shape of spectral features. Here we present the 3D spectra from our  0G, 3G, and 10G models near apastron and periastron. At periastron, the 10G spectra shows a lower flux than the 3G and 0G model, indicating a cooler earth-facing hemisphere. The 3D spectra show stronger spectral features than the 1D spectra, particularly for the CO$_{2}$ absorption feature present at  $\sim$4.2 microns due to the stronger temperature gradients present in the 3D model. 
    Near apastron, the lower temperatures cause the influence of our active drag to be smaller, resulting in similar spectra between the models shown.  }
    \label{fig: 3DCloudless_3Models}
\end{figure}

We additionally compare the cloudless 3D models to each other to judge the impact of magnetic field strength on low-resolution spectra, shown in  Figure \ref{fig: 3DCloudless_3Models}.
As with the 1-D spectra, we see the largest differences between models near periastron. The 0G and 3G models have very similar flux ratios, with the 3G being slightly cooler. 
When comparing the 0G and 10G spectra, we see a more significant difference at periastron, where the 0G spectra shows  higher flux ratios. Although their temperature structures are similar, there are regions in view at this phase where the 0G model is warmer (see Figures \ref{fig: TOI150b10Gvs0GNearPeri} and \ref{fig: TPProfNearPeriAll4Form2}). At apastron, the spectra are similar since our kinematic MHD prescription results in weaker drag at lower temperatures. If we compare these spectra to the 1D spectra shown in Figure \ref{fig: 1D CloudyCloudless}, we see the 3D spectra show a much stronger CO$_{2}$ absorption feature at near  $\sim$4.2 microns. This is due to the stronger temperature gradients present in the 3D models.  Additionally, there exists a water feature slightly redward of 6 microns that appears in emission for our 3D models due to atmospheric temperature inversions that appears strictly in absorption for the 1D models. This feature thus offers a testable prediction for 3D effects. At apastron, the 1D spectra has a higher flux ratio everywhere. This is unsurprising, as the 1D model represents a global average of the planet and assumes perfect heat redistribution while the the 3D models are ``diluted'' by the cooler profiles present on the entire visible hemisphere. 


\section{Discussion} \label{sec:Discussion}
TOI-150b is part of a growing population of eccentric hot Jupiters, and thus presents a unique opportunity to test the effects of strongly changing irradiation during the course of a single orbit. This results in multiple timescales including irradiation, radiative cooling, and our active drag to change in both magnitude and relative influence. This brings about temporal variability over the course of a single orbit, which was particularly strong in our 10G active drag model. 

\subsection{Implications for Observations} \label{subsec:implications}
This inter-orbit variability could be probed through spectroscopy. As shown in Section \ref{subsec: simspec}, the results of our simulations indicate variability in the spectra at differing magnetic field strengths. As magnetic field strength increases, the maximum net Doppler shift from cross-correlation decreases. The magnetic field strength of 30G has the lowest peak Doppler shift with a blueshift of only 0.25 km/s.   

This planet represents an excellent target for ground based high-resolution spectroscopy, which could probe the changing chemistry and temperature structure of the planet throughout its orbit. In particular, the emergence of a temperature inversion near perisastron could be directly detected and compared to apastron spectra, which we predict to have no temperature inversions. 

Differentiating between magnetic field strengths observationally for this planet may be difficult. The net Doppler shifts shown in Figure \ref{fig: AllModelsCC} differ by less than 0.25 km/s between magnetic models for most phases. While high-resolution spectroscopy currently does not have the precision to differentiate between levels of magnetic field strengths based on our estimates, it would be able to differentiate between the circular and eccentric cases. At low spectral resolution, there are differences between the 0G and 10G models near periastron, particularly in the strength of the CO$_{2}$ feature at 4.2 microns. 

Observing a phase curve of this planet would be informative as the eccentricity of the planet would be apparent through the day-night contrast and the hotspot offset. Additionally, the confirmation or lack thereof of secondary eclipse would allow for tighter constraints to be placed on the orbital dynamics of the system. The amplitude would additionally provide insight into the day-night contrast of the planet, allowing us to infer heating and cooling timescales for this planet. 


This object is also interesting from a planet formation perspective, as tidal forces are expected to circularize the orbit. The timescale this process should occur over is approximately $\approx3.46$ Gyr \citep{2006Adams, Kossakowski2019} for TOI-150b. Because this timescale is slightly larger than the age of the system, which was estimated by \cite{Kossakowski2019} to be $2.346^{+0.425}_{-0.901}$ Gyr, we expect that the planet is still in the process of circularizing.

\subsection{Comparison to Previous Work} \label{subsec:comparisons}
Our work is builds upon \cite{Kataria2013}, who studied the atmospheric dynamics of an eccentric hot Jupiter. Their models were substantially cooler than the ones presented here, with their hottest model having an equilibrium temperature of 1199K. They also did not include magnetic drag effects into their model. \cite{Kataria2013} does present a e=0.25 case, which is comparable to the e=0.26 models in this work. If we compare our 0G model to the models shown in \cite{Kataria2013}, similar atmospheric structures are present (as seen in Section \ref{sec:circvseccen}). These include the continued presence of a strengthened and narrowed equatorial super-rotating jet, westward motion at mid-latitudes, and an increased day-night temperature contrast. 

\cite{2021Mayorga} recently explored the effect of eccentricity in a 1D model of hot Jupiters. Their finding of an offset in the peak flux with periastron passage is further validated by our work. Their finding of a time-dependent thermal inversion is also present in this work, with the feature occurring near periastron and not near apastron, as expected. 


\cite{Beltz2021} saw that magnetic drag effects most greatly impact the upper atmosphere of the dayside of the planet, and that high levels of magnetic field strengths probed deeper into the atmosphere. We found the same results in our eccentric planet (Section \ref{subsec: mageffects}). Furthermore, \cite{Beltz2022a} investigated theoretical high-resolution emission spectra for hot Jupiter WASP-76b. We investigate the same type of spectra generated for TOI-150b in Section \ref{subsec: simspec}. Similar to our results, they found that the 0G model had a higher continuum flux than the models with magnetic drag. 
We also included a moderately strong magnetic field, which has shown to have the most inter-orbit variability. This strongly influences our spectra and, consequently, our cross-correlation results.

\subsection{Limitations} \label{subsec:limitsandfuture}

 Our kinematic MHD approach assumes that the planet's magnetic field is a dipole aligned with the axis of rotation. It is entirely possible that the planet's magnetic field is tilted or perhaps more complex, such as a quadrople. These different field orientations could impact the circulation pattern and affect the winds and Doppler shifts. However, there are no current constraints on any exoplanet magnetic field topology, making our aligned dipole assumption a reasonable place to start. In our solar system, this is an excellent approximation for Saturn--which has a nearly aligned dipole. Jupiter's dipole is slightly tilted (10$^\circ$) \citep{1993Russell}, but we leave the process of modeling tilted dipoles for future work. Thus, our approximation that the planet's magnetic field is a dipole aligned with the axis of rotation is reasonable.

 The 3D models of this planet assumed solar metallicity as an initial starting point. Should the planet deviate significantly from solar, the resulting temperature structure and spectra would be altered.

We additionally assumed in both the 1D and 3D models that the planet is rotating in a pseudo-synchronous rotation period, following the assumptions of \cite{Hut1981}. However, this rotation rate is unconstrained observationally. Different rotation rates would alter the circulation pattern and subsequent Doppler shifts measured from high-resolution spectra.

Our treatment of clouds in this work was limited, as they were only included in the 1D simulations. Cloud distribution could vary strongly spatially and temporally, which was not accounted for in this work. 

\subsection{Future Work}
Future observations are needed to test the predictions laid out in this paper, as there are currently no published spectra or phase curves of the planet. High-resolution observations at several phases would provide insight into the amount of broadening throughout the orbit, allowing for a direct probe of the atmospheric dynamics. Lower resolution \textit{JWST} spectra would provide insight into the presence and duration of an inverted atmosphere, which we predict to be present only during part of the planet's orbit.   Multiple observations spanning over many epochs  would be useful in characterizing the variability of the planet on different timescales.

Additionally, more theoretical work is needed to understand eccentric hot Jupiters and the temporal variability associated with them. Comparisons with other GCMs, including those with a more sophisticated radiative transfer routine such as correlated-k will determine the robustness of the features and predictions presented here. Given the expected variability of this planet, a more complex treatment of the atmospheric chemistry of the planet would be particularly informative, as the changing temperature structure could result in changing chemical abundances over the course of a single orbit. A 3D model coupled with clouds and hazes would provide a more detailed insight on expected cloud types and distributions.   Furthermore, future work should explore the effects of eccentricity and magnetic drag on transmission spectra across different resolutions, potentially with a more realistic treatment of MHD effects.


\section{Conclusions} \label{sec:Conclusion}
In this work, we present 3D models both with and without active magnetic drag of the planet TOI-150b, an eccentric gas giant with strong variation in its equilibrium temperature during the course of a single orbit. Our main conclusions are as follows:
\begin{itemize}
    \item The atmospheric structure of the planet changes significantly with the inclusion of eccentricity. We find the emergence and disappearance of temperature inversions, as expected from previous 1D models \citep{2021Mayorga}.
    \item We also see the emergence of significant westward motion at high latitudes. This westward motion is not found in our circular model, as expected from previous 3D GCMs of eccentric planets \citep{Kataria2013}. 
    \item The inclusion of kinematic MHD results in a magnetic circulation regime, particularly in the upper atmosphere, where flow in the hottest portion of the atmosphere is solely poleward. This effect is most prominent near periastron. The lower temperatures near apastron cause little difference in flow structure for the magnetic field strengths tested. 
    \item Increasing the magnetic field strength weakens the equatorial jet and allows the higher-latitude westward motion to extend deeper in the atmosphere and closer to the equator.
    \item Increasing magnetic field strength results in a decrease in net Doppler shifts in ground based high-resolution emission spectra. Moderate magnetic field strengths display inter-orbit variability in wind structure, leading to highly contrasting Doppler shifts when probing the same hemisphere of the planet from orbit to orbit.
    \item Eccentric hot Jupiters are a unique testbed to study atmospheric variability over a timescale of several days. We predict that the temperature structure, presence of a thermal inversion, and the phase-resolved net Doppler shifts show variability over the course of a single orbit, particularly for our models that include magnetic drag. 
    \item \textit{JWST} emission spectra over a broad wavelength range would provide a reference point to test our predictions. Specifically, measuring the strength of the 4.2 micron CO$_{2}$ feature and determining if the 6.2 micron H$_{2}$O feature is in emission or absorption would show evidence of 3D effects and potentially allow us to differentiate between our magnetic models.  
\end{itemize}

\section{Acknowledgments}
The authors would like to thank the reviewer for their helpful comments, which increased the quality of this work. This work was generously supported by a grant from the Heising-Simons Foundation as part of the 51 Pegasi b Fellowship Enhancements. Many of the calculations in this paper made use of the Great Lakes High Performance Computing Cluster maintained by the University of Michigan and the Zaratan cluster maintained by the University of Maryland. TDK acknowledges support from NASA XRP through grant number 80NSSC23K1172. 
J.R.L. acknowledges financial support from a Fluno Graduate Fellowship through the University of Wisconsin--Madison.

\bibliographystyle{aasjournal}
\bibliography{bib.bib}
\appendix
\section{A Note on Secondary Eclipse} \label{subsec: orbdyn}
Previously, \cite{Kossakowski2019} was unable to confirm the presence of a secondary eclipse of TOI-150b due to significant systematics in the data, notably setting an upper limit on the possible eclipse depth of 69$\pm$2 ppm. However, limitations in data quality, such as noise levels and the number of sectors of observations available, are not the only challenges in detecting a secondary eclipse. The specific angular orbital geometry of TOI-150b's orbit, influenced by its non-zero eccentricity, also plays an important role. While planets in circular orbits that transit their star's mid-plane typically undergo a secondary eclipse, those in eccentric orbits may not align similarly and thus may not allow a detection of a secondary eclipse.

Given TOI-150b's proximity to the JWST Continuous Viewing Zone, a secondary eclipse, if it occurs, would likely be detectable. 
To evaluate this possibility, we used a Monte Carlo method \citep{metropolis1949monte}. This method of generating probability distributions has been often applied to dynamical problems in exoplanets (see \citealp{Bonavita2012} and \citealp{Quinn2019}) to estimate the transit probability, drawing from Gaussian-distributed posterior parameters on TOI-150b's orbital geometry as published by \citet{Kossakowski2019}. We do this by sampling 100,000 sets of orbital elements from the published posteriors. For each drawn set of parameters, we evaluated whether a secondary eclipse would occur by modeling the planetary orbit and determining if it would pass directly behind the star within 1 $R_{*}$ of the stellar midplane from our line of sight. Our analysis finds only a 22\% chance that TOI-150b's orbit would exhibit a secondary eclipse geometrically detectable by JWST. SNR-based limitations may further decrease the feasibility of the measurement. 

\section{Additional GCM results}
\label{sec:Appendix}
\begin{figure*}
    \centering
    \includegraphics[width=6in]{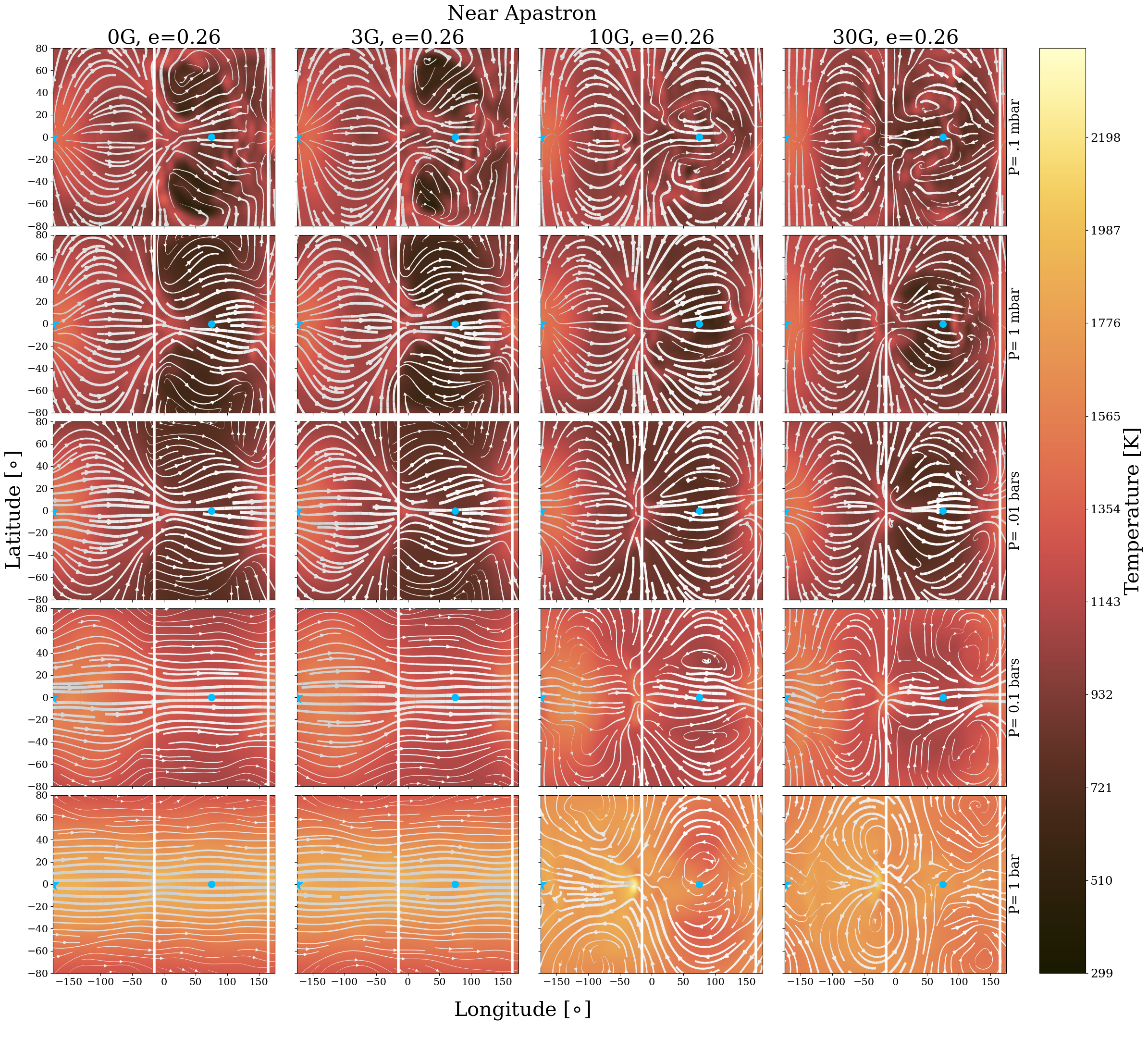}
    \caption{ Magnetic field strength along with orbital phase in a non-circular orbit can impact the wind and thermal structure of a hot Jupiter. Here we show the wind and thermal structure of TOI-150b near apastron at varying pressure levels (rows) and magnetic field strengths (columns). The Earth-facing hemisphere is between the two white solid lines, and the substellar point is denoted by the star (which at this phase, is near -180$^{\circ}$). The subobserver point is denoted by the circle. At the lower temperatures near apastron, the magnetic drag has a much smaller effect on the atmospheric structure than near periastron, and none of the cases show the typical behavior of magnetic circulation. However, the 10 and 30G cases do show slightly more uniform temperatures in the upper atmosphere.  
  } 
    \label{fig: StreamlineAll4NoDiff26}
\end{figure*}

In Figure \ref{fig: StreamlineAll4NoDiff26}, we present the eccentric models' thermal structures near apastron at varying pressure levels. This was done to see how the reduced level of stellar irradiation impacts the magnetic drag and subsequent temperature and winds of the planet. Since there is the lowest amount of stellar irradiation at this phase, the changes in the wind and thermal structures between the different magnetic field strengths are smaller than at periastron, as the magnetic drag timescale is much longer and hence much weaker at these lower temperatures. 

\begin{figure*}
    \centering
    \includegraphics[width=6in]{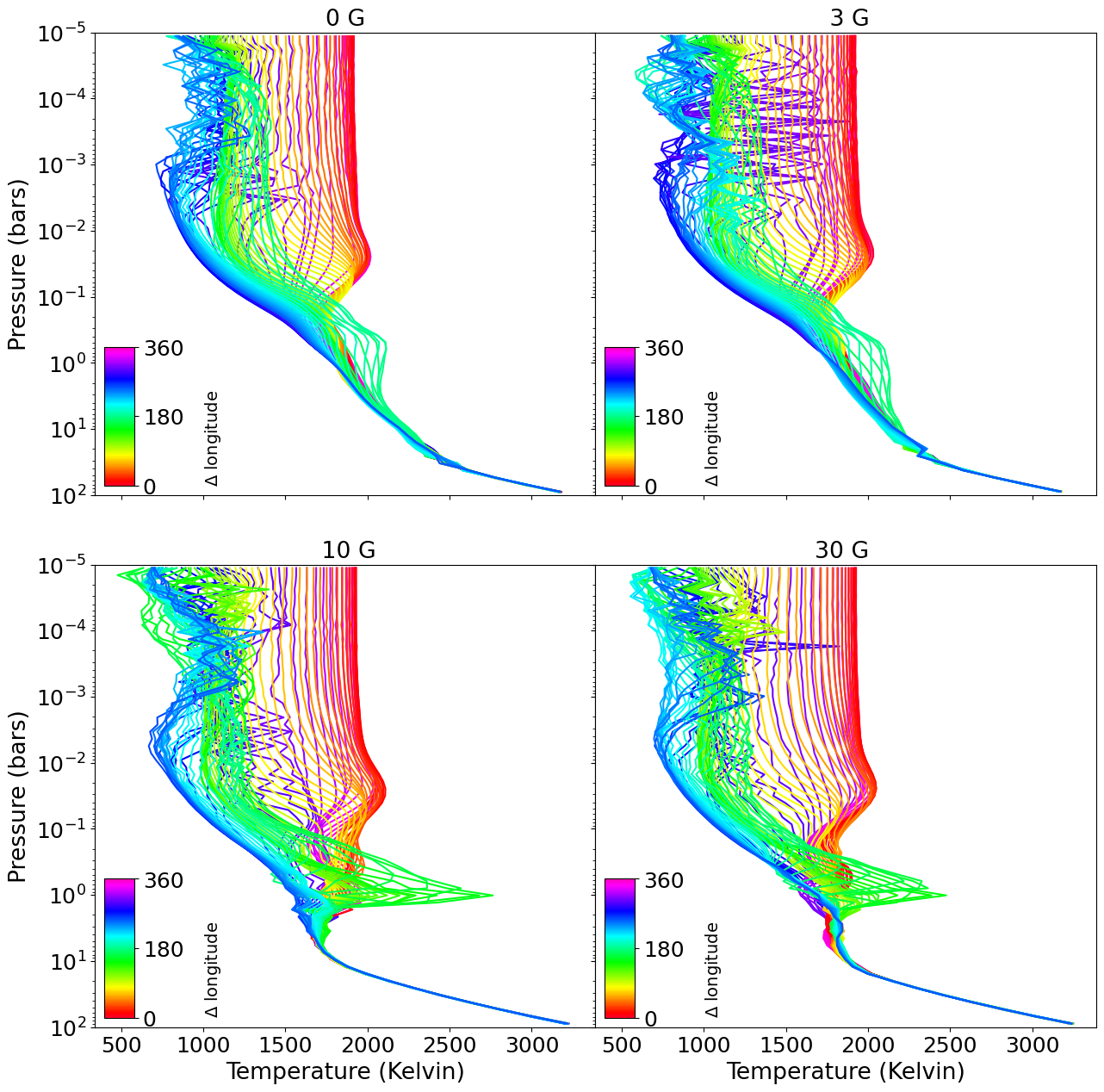}
    \caption{ The strength of the magnetic field applied to the planet's eccentric atmosphere can influence important thermal and pressure characteristics. Here we present snapshots of equatorial temperature-pressure profiles for each of our models with an eccentric orbit at a phase near periastron. The profiles are color-coded based on the distance to the current substellar longitude.  At this phase, each model shows a dayside inversion in the 0.1 to 0.01 bar range. The strength of this inversion increases as the field strength is increased, with the exception of the 10G case. In the 10G and 30G case, we see very strong inversions in the deep atmosphere. These inversions are due to the ohmic dissipation applied in the kinematic MHD scheme. } 
    \label{fig: TPProfNearPeriAll4Form2}
\end{figure*}

In Figure \ref{fig: TPProfNearPeriAll4Form2}, we display the equatorial temperature-pressure profiles, color-coded based on the difference in longitude from the substellar point,  for the eccentric models near periastron. As the dipole field strength is increased, the day night temperature contrast increases. The magnitude of the thermal inversion increases slightly as well, with the exception of the 10G case. The large inversions in the deep atmosphere for the 10G and 30G case are a result of increased Ohmic dissipation, which is deposited deep in the atmosphere as part of our kinematic MHD scheme. 

\begin{figure}
    \centering
    \includegraphics[width=6in]{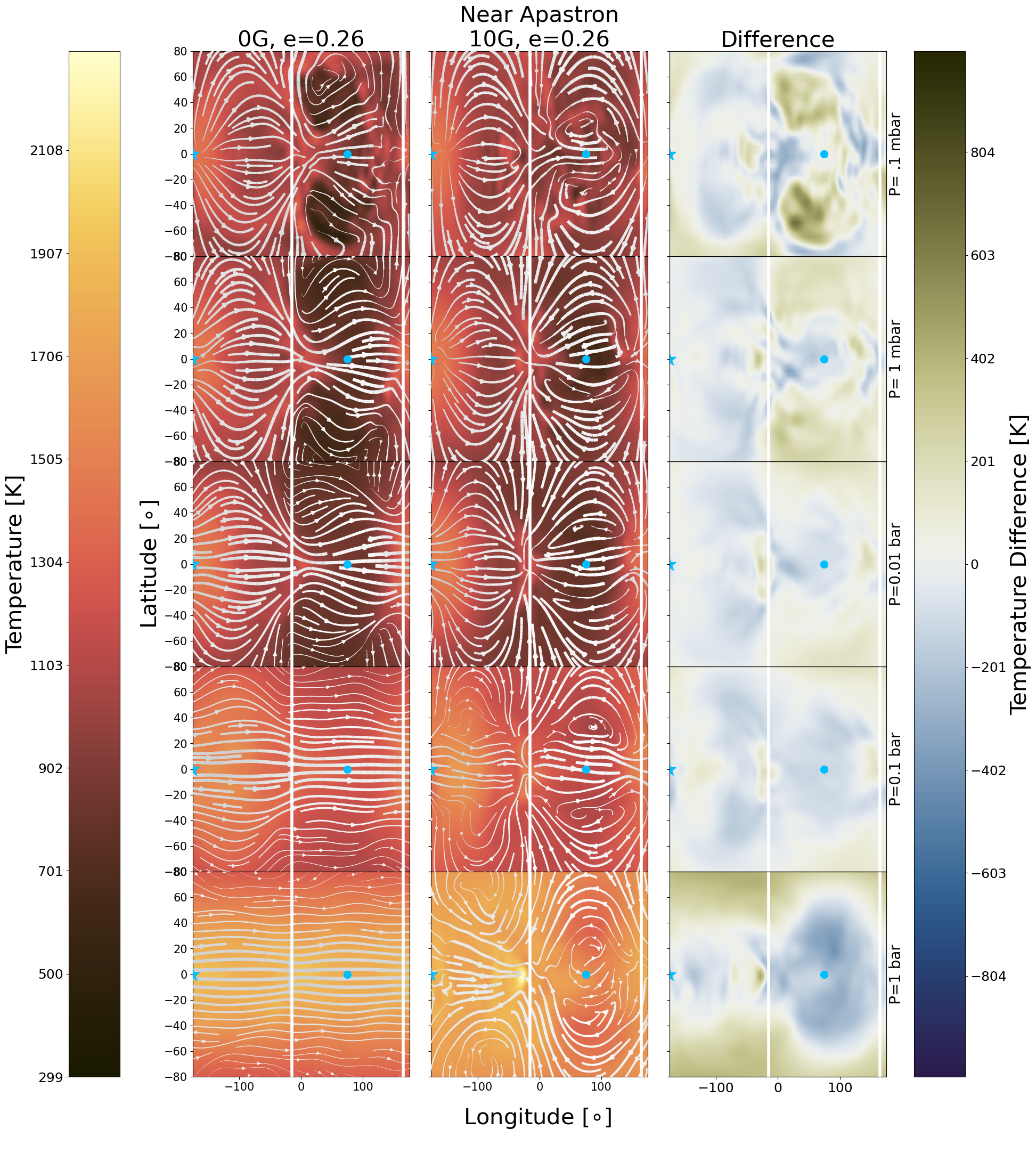}
    \caption{ Similar to Figure \ref{fig: TOI150b10Gvs0GNearPeri}, but showing temperature and streamlines at an orbital phase near apastron. The Earth-facing hemisphere is between the white solid lines, and the substellar point is denoted by the star (which at this phase, is near -180$^{\circ}$, on the far left of each subplot). The subobserver point is denoted by the circle. At lower pressure levels, the 10G case is hotter at the anti-stellar and sub-stellar points, whereas the 0G case is hotter elsewhere. However, as we probe deeper into the atmosphere, we see that the 0G case is hotter at the substellar and anti-stellar longitudes at depth, whereas the 10G case is hotter elsewhere.   } 
    \label{fig: TOI150b10Gvs0GNearApas}
\end{figure}

In Figure \ref{fig: TOI150b10Gvs0GNearApas}, we show the difference between the 0G eccentric model and the 10G eccentric model near apastron, where stellar irradiation is the weakest. As previously discussed, the decrease in stellar irradiation results in cooler atmospheric temperatures and reduces the influence of our active magnetic drag. The differences in temperature between the 0G and 10G models are smaller in magnitude here than near periastron (see Figure \ref{fig: TOI150b10Gvs0GNearPeri}). The lack of a jet in the deep atmosphere of the 10G model is still apparent.  

\end{document}